\documentclass{pasj01}

\usepackage{lscape} 

\begin{document}
\SetRunningHead{Author(s) in page-head}{Running Head}

\title{Q-type asteroids: Possibility of non-fresh weathered surfaces}

\author{%
   Sunao \textsc{Hasegawa},\altaffilmark{1,*}
   Takahiro \textsc{Hiroi},\altaffilmark{2}
   Katsuhito \textsc{Ohtsuka},\altaffilmark{3}
   Masateru \textsc{Ishiguro},\altaffilmark{4}
   Daisuke \textsc{Kuroda},\altaffilmark{5}
   Takashi \textsc{Ito},\altaffilmark{6}
   and
   Sho \textsc{Sasaki}\altaffilmark{7}
}
 \altaffiltext{1}{Institute of Space and Astronautical Science, Japan Aerospace Exploration Agency, 3-1-1 Yoshinodai, Chuo-ku, Sagamihara 252-5210, Japan}
 \email{hasehase@isas.jaxa.jp}
 \altaffiltext{2}{Department of Earth, Environmental and Planetary Sciences, Brown University, Providence, Rhode Island 02912, USA}
 \altaffiltext{3}{Tokyo Meteor Network, 1-27-5 Daisawa, Setagaya-ku, Tokyo 155-0032, Japan}
 \altaffiltext{4}{Department of Physical and Astronomy, Seoul National University, Gwanak-ro, Gwanak-gu, Seoul 08826, Korea}
 \altaffiltext{5}{Okayama Astronomical Observatory, Kyoto University, 3037-5 Honjo, Kamogata-cho, Asakuchi, Okayama 719-0232, Japan}
 \altaffiltext{6}{Center for Computational Astrophysics, National Astronomical Observatory of Japan, 2-21-1 Osawa, Mitaka, Tokyo 181-8588, Japan}
 \altaffiltext{7}{Department of Earth and Space Science, Osaka University, 1-1 Kaneyama-cho, Toyonaka, 560-0043, Japan}


\maketitle

\begin{abstract}
Itokawa particles, which are the recovered samples from the S-complex asteroid 25143 Itokawa by the Hayabusa spacecraft, demonstrate that S-complex asteroids are parent bodies of ordinary chondrite meteorites.
Furthermore, they clarify that the space weathering age of the Itokawa surface is of the order of several thousand years.
Traditionally, Q-type asteroids have been considered fresh-surfaced.
However, as the space weathering timescale is approximately three orders of magnitude lesser than the conventionally considered age, the previously proposed formation mechanisms of Q-type asteroids cannot sufficiently explain the surface refreshening.
In this study, we propose a new hypothesis on the surface state of Q-type asteroids: Q-type asteroids have a non-fresh weathered surface with a paucity of fine particles.
For verifying this hypothesis, laboratory experiments on the space weathering of ordinary chondrites are performed. 
Based on the results of these experiments, we found that large (more than 100 \micron) ordinary chondritic particles with space weathering exhibit spectra consistent with Q-type asteroids.
\end{abstract}

\section{Introduction}
Over the past 40 years, almost no main-belt asteroids having the characteristics of ordinary chondritic surfaces have been discovered, which was a major conundrum.

Using spectrophotometry in the visible region, it was discovered that the surface of the asteroid 4 Vesta was similar to those of HED meteorites \citep{McCord1970}.
Subsequently, early spectrophotometric surveys of other types of asteroids were performed (e.g., \cite{Chapman1973a}; \cite{Chapman1973b}).
Although certain near-earth asteroids were detected with the characteristics of ordinary chondrites, which have been observed in most recovered meteorites on earth \citep{Mason1962}, such characteristics were not found in main-belt asteroids (\cite{Chapman1973C}; \cite{Egan1973}; \cite{McCord1974}; \cite{McFadden1985}).
Instead of asteroids composed of ordinary chondrites, also called Q-type asteroids \citep{Tholen1984}, many S-complex asteroids with a reddish gradient in the spectra, compared to ordinary chondrites, were detected \citep{Chapman1975}.
Later, large-scale spectrophotometry and spectroscopic surveys for asteroids have been conducted since (\cite{Chapman1979}; \cite{Zellner1985}; \cite{Bus2002}; \cite{Lazzaro2004}; \cite{Carvano2010}).
However, S-complex asteroids rather than ordinary chondritic-like asteroids are found to be the major constituents of the main belt.

For solving this dilemma on the difference in color between S-complex asteroids and ordinary chondrites, two different solutions were suggested.
One was that S-complex asteroids were covered by a surface layer of stony-iron meteorites, suggesting that the ordinary chondrites originated from a specific limited asteroidal area (\authorcite{Gaffey1984} \yearcite{Gaffey1984}, \yearcite{Gaffey1986}; \cite{Fanale1992}; \authorcite{Gaffey1993} \yearcite{Gaffey1993}, \yearcite{Gaffey1998}); the other was that the color of S-complex asteroids was altered from those of ordinary chondrites by the `space weathering' process (\cite{Feierberg1982}; \cite{McFadden1983}; \cite{Pieters1984}).

It was first considered that there was `space weathering' as ray degradation process because  the only lunar newer craters have rays \citep{Gold1955}.
\citet{Hapke1962} and \citet{Wehner1963} suggested that the darkness on the moon was at least partially caused by solar wind.
\citet{Hapke1965} proposed that proton irradiation on igneous powder rocks can produce photometric, polarimetric, and visible spectroscopic properties similar to the lunar surface.
\citet{Hapke1975} also showed that not only solar wind sputtering but also impact vaporization contributed to material darkening in the lunar regolith.
The discovery of rims on lunar soil grain samples with iron nanoparticles, obtained by the Apollo missions, proved the space weathering process on the moon (\authorcite{Keller1993} \yearcite{Keller1993}, \yearcite{Keller1997}).

However, the idea of the space weathering process as the solution to the S-complex asteroid problem was not considered superior to that of the origin of stony-iron meteorites (e.g., \cite{Gaffey1993a}).
\citet{Hiroi1993} demonstrated that the reflectance spectra of S-complex asteroids can be explained by the stony-iron model, but also concurrently implied that the model for S-complex asteroids may not be appropriate.
\citet{Clark1994} argued that the reddish slope continuum of S-complex asteroids cannot be produced by metallic iron alone.
\citet{Binzel1996} demonstrated that S-complex near-earth asteroids have a non-discrete spectra between those of Q-type asteroids and S-complex asteroids, through ground-based observations, and the results indicated that the surfaces of S-complex asteroids were altered by the time-dependent spaceweathering process.
The results of the {\it in-situ} observations of S-complex asteroids by the Galileo, NEAR-Shoemaker, and Hayabusa spacecraft indicated that the space weathering process caused asteroid-surface reddening (\cite{Chapman1996}; \cite{Trombka2000}; \cite{Veverka2000}; \cite{Hiroi2006}).
A mixture of rare ordinary chondrites including melted, gas-rich, and shocked were considered (\cite{Clark1992}; \cite{Britt1994}) to cause space weathering, but this could not reproduce the reddish-slope spectra experimentally.
Laboratory experiments using pulsed laser to simulate micrometeorite impacts was attempted by \citet{Moroz1996} and was later improved to reproduce the optical-property alteration of space weathering, namely, spectral reddening as well as the darkening of iron-bearing silicates \citep{Yamada1999}.
Ion ($\mathrm{He^{+}}$) irradiation simulating solar wind bombardment produced nanophase iron on silicate surface \citep{Dukes1999}.
The discovery of iron nanoparticles on pulse-laser irradiated samples strongly indicated that the spectra of S-complex asteroids were altered by the space weathering process \citep{Sasaki2001}.
Returned samples by the Hayabusa spacecraft proved that the S-complex asteroid 25143 Itokawa was composed of ordinary chondrite material LL5/6, (\cite{Nakamura2011}; \cite{Tsuchiyama2011}; \cite{Yurimoto2011}) and that nanophase iron particles that cause spectral reddening/darkening were produced on the Itokawa \citep{Noguchi2011}.

Although these Itokawa particles established the origin of S-complex asteroids, solving the conundrum for 40 years, it is important to note that they created a new enigma on the space weathering process, as described below. 

Conventionally, the time scale of space weathering is considered to be $\sim$1 Myr based on the results of laboratory experiments and spectroscopic observations of S-complex asteroid families (\cite{Vernazza2009}; \cite{Marchi2012}).
However, the following were determined based on three different analyzes of the Itokawa particles.

\begin{description}
\item{$\bullet$}
Based on the concentration of the Neon ion and the number density of the solar flare tracks of Itokawa particles, it was estimated that the particle size remained within few centimeters for a duration of 150--550 yr \citep{Nagao2011} and approximately thousands to tens of thousands of years (\cite{Noguchi2014}; \cite{Keller2014}), respectively.
\item{$\bullet$}
\citet{Matsumoto2018} demonstrated that the dominant cratering process on the Itokawa particles in the submicrometer range was caused by the secondary ejecta created by primary impacts on the Itokawa surface.
If it is assumed that the secondary crater formation rate is equivalent to the lunar impact flux, the measured age based on the number of submicrometer craters can be estimated to be approximately thousands of years.
\item{$\bullet$}
\citet{Bonal2015} showed that the spectra of Itokawa particles measured by visible-IR and Raman microspectroscopy were the same or as reddish as the spectra of the Itokawa obtained by ground-based telescopes.
This result indicates that the Itokawa particles brought back by the Hayabusa represented the regolith particles on the Itokawa surface.

In addition, they estimated that the space weathering age of the Itokawa particles was several Myr by extrapolating the results of laboratory simulation.
However, this result casts doubts on the dating method based on the laboratory simulation of space weathering.
\end{description}
A combination of these results using cosmochemical, crater counting, and spectral analysis of the Itokawa particles indicates that the space weathering age of the Itokawa surface is in the order of approximately thousands of years.

The asteroid surface matures from a refreshing ordinary chondritic surface to an S-complex asteroid-like surface due to the space weathering process.
As this is an irreversible process, several models have been proposed for creating a Q-type asteroid-like surface.
A model was proposed in which the regolith on the S-complex asteroid surface was peeled-off during close encounters with a planet, and the unweathered subsurface material was exposed to become a Q-type asteroid (\cite{Binzel2010}; \cite{Nesvorny2010}; \cite{DeMeo2014}).
In another model, the Yarkovsky-O'Keefe-Radzievskii-Paddack (YORP) effect was considered to play a role in accelerating the rotation of the asteroid and removing the weathered regolith on the asteroid surface to expose the Q-type asteroid-like surface (\cite{Polishook2014}; \cite{Graves2018}).
\citet{Delbo2014} proposed that refreshed regoliths were formed by thermal fatigue induced by diurnal temperature variations.
However, the time scale required for these formation models is more than several million years except three asteroid pairs in the main belt: 6070 Rheinland -- 54827 Kurpfalz, 51609 2001 $\mathrm{HZ_{32}}$ -- 322672 1999 $\mathrm{TE_{221}}$, and 19289 1996 $\mathrm{HY_{12}}$ -- 278067 2006 $\mathrm{YY_{40}}$ to have probable ages of 16 Kyr, $\sim$540 kyr, and \textgreater 630 kyr, respectively (\cite{Polishook2014}; \cite{Pravec2019}), and if the time scale of space weathering is around thousands years, it is not so easy to explain the surface formation of Q-type asteroids with these models.

It was found that the surfaces of asteroids greater than km-size were covered with regolith, using spacecrafts such as the Dawn, Galileo, NEAR-Shoemaker, and Rosetta (1 Ceres: \cite{Prettyman2017}; 4 Vesta: \cite{Jaumann2012}; 21 Lutetia: \cite{Sierks2011}; 243 Ida: \cite{Belton1994}; 253 Mathilde: \cite{Clark1999}; 433 Eros \cite{Yeomans2000}; 951 Gaspra: \cite{Belton1992}; 2867 Steins: \cite{Leyrat2011}).
On the contrary, it was discovered that the surfaces of asteroids less than 1-km in diameter were dense with boulders by the Hayabusa, Hayabusa2, and OSIRIS-REx spacecrafts (25143 Itokawa: \cite{Fujiwara2006}; 101955 Bennu: \cite{DellaGiustina2019}; 162173 Ryugu: \cite{Sugita2019}).
\citet{Ishiguro2017} showed that the surface of the Q-type asteroid 1566 Icarus with an equivalent diameter of 1.4 km \citep{Greenberg2017} was covered by coarse and/or medium sand with a diameter of several hundreds of micrometers, through polarimetric observations.
Note that the sizes of the known Q-type asteroids are almost equal to that of the Icarus or smaller.

Generally, meteorite spectrum becomes bluish with the increase in particle size (e.g., \cite{Johnson1973}; \cite{Miyamoto1981}; \cite{Binzel2015}; \cite{Reddy2016}; \cite{Vernazza2016}).
If the particle size is very large, the slope of the spectrum may be negative.
On the other hand, the space weathering process progresses with the reddening of the meteorite spectrum. 

Combining these facts, we propose a new model in which Q-type asteroids are composed of large particles undergoing the space weathering process.
The subsequent sections of this paper describe the calculation methods and results for confirming the proximity encounters between Q-type asteroids and massive objects (section 2), propose methods for the laboratory simulation of space weathering and the measurement of the sample spectra, and present the obtained spectral results with changes in the particle sizes and weathering degree (section 3), discuss the implications of our findings (section 4), and summarize the study (section 5).

\section{Orbital evolution analysis}
\subsection{Methods}
For the conventional resurface model of Q-type asteroids, an investigation of close encounters with massive objects during the weathering process are crucial.
For evaluating whether such close encounters cause surface renewal, trajectory-evolution calculation utilizes the backward numerical integration of the post-Newtonian equation of motion.
The calculation code uses the SOLEX package based on the Bulirsh-Stoer integrator \citep{Vitagliano1997}.
The coordinates and velocities of massive objects such as perturbers are calculated using the JPL Planetary and Lunar Ephemeris DE/LE431, which includes the eight planets, the Moon, two dwarf planets Pluto and Ceres, and two large main-belt asteroids Pallas and Vesta.
The SOLEX can accurately process close encounters through a routine that performs automatic time-step adjustments.

The list of Q-type asteroids for orbit history is mainly quoted from \citet{Binzel2019} and \citet{Hasegawa2018}, \citet{Lin2018}, \citet{Perna2018}, and \citet{Popescu2019} to a lesser extent.
Of the 185 Q-type asteroids, 143 are numbered asteroids and 42 are unnumbered. 
As the orbit parameters of certain unnumbered asteroids involve considerable errors and cannot be used, the orbit calculation of only 18 (out of 42) unnumbered asteroids, whose JPL condition code was zero (orbit uncertainty estimate 0--9, 0: good and 9: highly uncertain), was performed.

\citet{Marchi2006} suggested a criterion that resurface occurs when the asteroid approaches a massive object, within twice the Roche radii.
In this study, the Marchi's resurface radii was utilized as a criterion for massive surface refreshening. 

As the Q-type-asteroid density for the resurface criterion, 1000 kg $\mathrm{m^{-3}}$, which is the lower limit of the density of S-complex asteroids \citep{Carry2012}, was utilized.
The lower density limit was set because high-porosity asteroids are assumed.

\subsection{Orbit history results}
\begin{figure*}
  \begin{center}
    \FigureFile(80mm,80mm){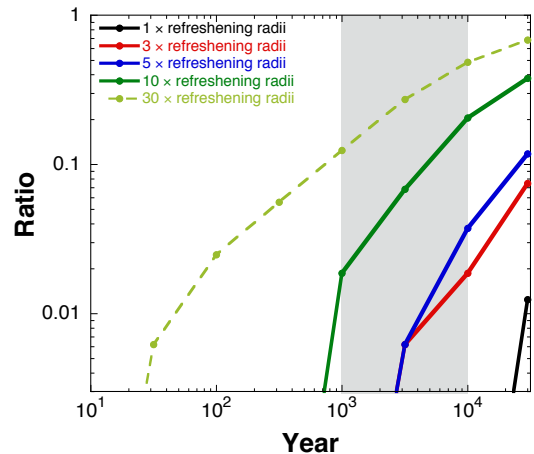}
  \end{center}
  \caption{Probability of surface refreshening for Q-type asteroids by planetary encounter during the space weathering process.
On the time scale of space weathering (indicated by the gray box), all the Q-type asteroids are not sufficiently close to the planets for surface refreshening.

}
\label{fig:Orbit}
\end{figure*}

As per the numerical orbit propagation, close encounters that affect the refreshening criterion occur against four inner planets located among massive objects.
Figure \ref{fig:Orbit} shows the time evolution of occurrence rate that the Q-type asteroids approaching the planets.
If Marchi's resurfacing radii are considered as the resurface criterion, there are no refreshes caused by close encounters with planets within $\mathrm{10^{3}}$--$\mathrm{10^{4}}$ yr.
On extrapolating this result to a time scale of 10 Myr from 30 kyr, the results are consistent with the numerical estimate employing the particle-in-a-box computations of \citet{Marchi2006}.

Using the orbit history of asteroid 25143 Itokawa, the upper limit of the refreshening due to close encounters with planets is estimated.
As Itokawa has a stable orbit for the past 200 years \citep{Yoshikawa2006}, its trajectory can be traced back to that point.
Itokawa approached the earth at a minimum distance of 28 times Marchi's refreshening radii approximately 200 years ago.
However, it is clear that its surface has not been refreshed.
Therefore, it can be considered that surface renewal does not occur even if the asteroid approaches to 30 times Marchi's refreshening radii.
Even if it is overestimated, the close planetary encounters of nearly half the Q-type asteroids can be explained; however, the other half remain unexplained. 

Consequently, as the surfaces of Q-type asteroids cannot be refreshed at the frequency of the close encounters calculated thus far, it is necessary to consider other mechanisms.

\section{Reflectance spectroscopy}
The spectral reflectance of ordinary chondrites with controlled particle size and degree of space weathering were obtained for verifying the reproduced spectra of Q-type asteroids under the space weathering condition.

\citet{Miyamoto1981} demonstrated that the spectral changes in ordinary chondrites depend upon the particle size. 
However, their spectral measurements of large particles included small particles as well, and the effect of small particle sizes was not eliminated in the spectral measurement of large particles.
In this study, particles of three sizes were prepared: chips, 125--500 \micron, and \textless125 \micron\ samples.

\subsection{Space weathering experiments}
According to the method pioneered by \citet{Yamada1999}, 7-ns pulse-laser irradiation experiments were performed on the chip and pellet samples of ordinary chondrites.
As powder can be blown-off by laser irradiation, pellets were prepared instead of powder samples.
Each pellet sample was made by pressing a powder sample of less than 125 \micron\ in a copper dish with an inner-diameter of 12 mm and outer-rim of 2--4 mm at a pressure of 30--40 MPa for 1 min.
For each chip sample, a relatively flat naturally broken surface was selected; the irradiated area is approximately 8 mm in diameter.
The applied accumulated total laser energy was 5--80 mJ, depending upon the sample. 
The vacuum pressure was typically $\mathrm{10^{-3}}$ Pa.

\subsection{Spectral reflectance measurements}
The ultraviolet to near-infrared reflectance spectra of the chip, powder, and pellet samples were measured using a bidirectional spectrometer at the Reflectance Experiment Laboratory (RELAB) in Brown University \citep{Pieters2004} or the Bunko-Keiki DRS-25 spectrometer at the University of Tokyo or Mizusawa VLBI Observatory of the National Astronomical Observatory of Japan.
Such spectrometers produce highly comparable reflectance spectra, according to an interlaboratory comparison study \citep{Hiroi2016}.
The main differences between these spectrometers were that the RELAB spectrometer had each sample spun at a speed of 1.5 s per rotation, and had considerably narrower incidence and emergence light cone angles, compared to the Bunko-Keiki spectrometer.

The incidence and emergence angles were set to 30 and 0 degrees, respectively, as the standard viewing geometry, and the data were collected over a wavelength range of either 0.3--2.6 \micron\ or 0.25--2.5 \micron\ at 5 nm intervals.
The spectral data over the common wavelength range (0.3--2.5 \micron) between the two spectrometers were used in this study. 
Pressed halon or Spectralon from Labsphere Inc. (North Sutton, NH) was used as the standard material, and raw reflectance spectra of our samples relative to one of these standards were corrected for the reflectance values and absorption bands (notable beyond 2 \micron\ in wavelength) using the RELAB calibration table for the respective standards.
The incident beam size was approximately 3--7 mm in diameter for the RELAB spectra depending upon the sample size, and approximately 2 $\times$ 3 mm for the Bunko-Keiki spectrometer.

All the spectral data, including those measured by the Bunko-Keiki spectrometer, were added to the RELAB database\footnotemark[1].
These spectral data are either already in the public domain or will be released once this paper is published.
\footnotetext[1]{http://www.planetary.brown.edu/relab/}

\subsection{Results of spectroscopy for ordinary chondrites}
The spectra of 21 ordinary chondrite meteorites were obtained under each of the three surface conditions: chip, 125--500 \micron, and \textless125 \micron.
Space weathering experiments were conducted on 12 out of the 21 samples.
Table \ref{tab:M} lists the information on the measured ordinary chondrites in this study.

For studying Q-type asteroids, it is necessary to remove samples with spectra that are not unique to ordinary chondrites. 
For this purpose, the meteorites were classified using the Bus-DeMeo taxonomy (\cite{DeMeo2009}; \cite{Binzel2019}), and a taxonomic tool available on the world wide web\footnotemark[2] was utilized.
The classified results of the meteorites as per the Bus-DeMeo taxonomy are also shown in table \ref{tab:M}.
Based on the classification results, in this study, spectral comparisons were performed for 15 meteorites.
\footnotetext[2]{http://smass.mit.edu/minus.html}

\begin{longtable}{llllllll}
  \caption{Spectroscopic circumstances of the ordinary chondrites.}\label{tab:M}
  \hline
    Meteorite & Meteorite & Physical & Laser & RELAB     & RELAB         & Bus-DeMeo & Slope of \\ 
    name      & type      & form     & int. [mJ]   & sample ID & spectral file & taxonomy  & B-D tax  \\
\endfirsthead
  \hline
    Name      & Type      & form     & Int. [mJ]   & Sample ID & Spectral file & B-D tax   & Slope    \\
  \hline
\endhead
  \hline
\endfoot
  \hline
\multicolumn{4}{@{}l@{}}{\hbox to 0pt{\parbox{200mm}{\footnotesize
\footnotemark[$*$] These meteorites are not used because their spectral type is unique (see section 3.3).\\
\footnotemark[$\dagger$] These laser irradiated samples are not used because the spectrum slope of the pellet before laser irradiation is bluer than those of the chip (see section 3.3).
}}}
\endlastfoot
  \hline
Cherokee Springs&LL6&Chip&0&OC-TXH-001-A&C1OC01A&O&$-$0.0551\\
&&125--500 \micron&0&OC-TXH-001-B&C1OC01B&O&$-$0.0110\\
&&\textless125 \micron&0&OC-TXH-001-C&C1OC01C&O&0.0999\\
  \hline
Hamlet\footnotemark[$*$]&LL4&Chip&0&OC-TXH-002-A&C1OC02A&K&$-$0.1026\\
\footnotemark[$*$]&&Chip&20&OC-TXH-002-A20&C1OC02A20&S&0.0842\\
\footnotemark[$*$]&&Chip&40&OC-TXH-002-A40&C1OC02A40&S&0.1344\\
\footnotemark[$*$]&&Chip&60&OC-TXH-002-A60&C1OC02A60&S&0.1622\\
\footnotemark[$*$]&&125--500 \micron&0&OC-TXH-002-B&C1OC02B&S&0.1641\\
\footnotemark[$*$]&&\textless125 \micron&0&OC-TXH-002-C&C1OC02C&Sq&0.145\\
  \hline
Harleton&L6&Chip&0&OC-TXH-003-A&C1OC03A&Q&$-$0.0547\\
&&125--500 \micron&0&OC-TXH-003-B&C1OC03B&O&$-$0.0308\\
&&\textless125 \micron&0&OC-TXH-003-C&C1OC03C&Q&0.0796\\
  \hline
Mezo-Madaras\footnotemark[$*$]&L3.7&Chip&0&OC-TXH-004-A&C1OC04A&Sq&$-$0.1084\\
\footnotemark[$*$]&&125--500 \micron&0&OC-TXH-004-B&C1OC04B&Sq&$-$0.0610\\
\footnotemark[$*$]&&\textless125 \micron&0&OC-TXH-004-C&C1OC04C&Sq&0.0615\\
  \hline
Monroe\footnotemark[$*$]&H4&Chip&0&OC-TXH-005-A&C1OC05A&Q&$-$0.1183\\
\footnotemark[$*$]&&125--500 \micron&0&OC-TXH-005-B&C1OC05B&Sq&$-$0.0965\\
\footnotemark[$*$]&&\textless125 \micron&0&OC-TXH-005-C&C1OC05C&Sq&0.0343\\
  \hline
Ehole &H5&Chip&0&OC-TXH-006-A&C1OC06A&O&$-$0.0254\\
&&Chip&20&OC-TXH-006-A20&C1OC06A20&Q&0.0756\\
&&Chip&40&OC-TXH-006-A40&C1OC06A40&Q&0.0576\\
&&125--500 \micron&0&OC-TXH-006-B&C1OC06B&O&0.0255\\
&&\textless125 \micron&0&OC-TXH-006-C&C1OC06C&O&0.0208\\
  \hline
Paragould\footnotemark[$*$]&LL5&Chip&0&OC-TXH-007-A&C1OC07A&Ch&0.0023\\
\footnotemark[$*$]&&Chip&20&OC-TXH-007-A20&C1OC07A20&Ch&0.0013\\
\footnotemark[$*$]&&Chip&40&OC-TXH-007-A40&C1OC07A40&Xc&0.0697\\
\footnotemark[$*$]&&Chip&60&OC-TXH-007-A60&C1OC07A60&Xn&0.0612\\
\footnotemark[$*$]&&125--500 \micron&0&OC-TXH-007-B&C1OC07B&B&$-$0.0009\\
\footnotemark[$*$]&&\textless125 \micron&0&OC-TXH-007-C&C1OC07C&Q&0.0435\\
\footnotemark[$*$]&&Chip&0&OC-TXH-007-D&C1OC07P&B&$-$0.0241\\
\footnotemark[$*$]&&\textless125 \micron\ pellet&0&OC-TXH-007-P&C1OC07P05&B&$-$0.1761\\
\footnotemark[$*$]&&\textless125 \micron\ pellet&5&OC-TXH-007-P05&C1OC07P15&B&$-$0.0993\\
\footnotemark[$*$]&&\textless125 \micron\ pellet&15&OC-TXH-007-P15&C1OC07D&Xn&0.0271\\
  \hline
Ochansk&H4&Chip&0&OC-TXH-008-A&C1OC08A&Sq&$-$0.1133\\
&&125--500 \micron&0&OC-TXH-008-B&C1OC08B&Q&$-$0.1196\\
&&\textless125 \micron&0&OC-TXH-008-C&C1OC08C&Q&$-$0.0284\\
  \hline
Olivenza&LL5&Chip&0&OC-TXH-009-A&C1OC09A&Q&$-$0.1084\\
&&125--500 \micron&0&OC-TXH-009-B&C1OC09B&O&$-$0.0643\\
&&\textless125 \micron&0&OC-TXH-009-C&C1OC09C&O&0.0338\\
  \hline
Alta'ameem&LL5&Chip&0&OC-TXH-010-A&C1OC10A&Q&$-$0.0918\\
&&125--500 \micron&0&OC-TXH-010-B&C1OC10B&Q&$-$0.0606\\
&&\textless125 \micron&0&OC-TXH-010-C&C1OC10C&Q&0.0174\\
  \hline
Chateau Renard&L6&Chip&0&OC-TXH-011-A&C1OC11A&Q&$-$0.0881\\
&&Chip&20&OC-TXH-011-A20&C1OC11A20&Q&0.0789\\
&&Chip&40&OC-TXH-011-A40&C1OC11A40&Q&0.1327\\
&&Chip&60&OC-TXH-011-A60&C1OC11A60&Q&0.1639\\
&&Chip&80&OC-TXH-011-A80&C1OC11A80&Q&0.1769\\
&&125--500 \micron&0&OC-TXH-011-B&C1OC11B&O&$-$0.0572\\
&&\textless125 \micron&0&OC-TXH-011-C&C1OC11C&Q&0.0604\\
&&\textless125 \micron\ pellet&0&OC-TXH-011-D&C1OC11D&Q&$-$0.0820\\
&&\textless125 \micron\ pellet&5&OC-TXH-011-D05&C1OC11D05&Q&0.018\\
&&\textless125 \micron\ pellet&15&OC-TXH-011-D15&C1OC11D15&Sq&0.1375\\
&&\textless125 \micron\ pellet&35&OC-TXH-011-D35&C1OC11D35&Sq&0.2429\\
  \hline
Appley Bridge&LL6&Chip&0&OC-TXH-012-A&C1OC12A&Q&$-$0.1105\\
&&Chip&20&OC-TXH-012-A20&C1OC12A20&Q&0.078\\
&&Chip&40&OC-TXH-012-A40&C1OC12A40&Q&0.1476\\
&&125--500 \micron&0&OC-TXH-012-B&C1OC12B&Q&$-$0.0556\\
&&\textless125 \micron&0&OC-TXH-012-C&C1OC12C&Q&0.0569\\
&&\textless125 \micron\ pellet&0&OC-TXH-012-P&C1OC12P&Q&$-$0.1702\\
\footnotemark[$\dagger$]&&\textless125 \micron\ pellet&5&OC-TXH-012-P05&C1OC12P05&Q&$-$0.1225\\
\footnotemark[$\dagger$]&&\textless125 \micron\ pellet&15&OC-TXH-012-P15&C1OC12P15&Sq&$-$0.0008\\
\footnotemark[$\dagger$]&&\textless125 \micron\ pellet&35&OC-TXH-012-P35&C1OC12P35&Sq&0.1149\\
  \hline
Athens&LL6&Chip&0&OC-TXH-013-A&C1OC13A&Q&$-$0.0719\\
&&125--500 \micron&0&OC-TXH-013-B&C1OC13B&Q&$-$0.0557\\
&&\textless125 \micron&0&OC-TXH-013-C&C1OC13C&Q&0.0399\\
&&\textless125 \micron\ pellet&0&OC-TXH-013-P&C1OC13P&Q&$-$0.2012\\
\footnotemark[$\dagger$]&&\textless125 \micron\ pellet&5&OC-TXH-013-P05&C1OC13P05&S&$-$0.1244\\
\footnotemark[$\dagger$]&&\textless125 \micron\ pellet&15&OC-TXH-013-P15&C1OC13P15&S&$-$0.0353\\
  \hline
Chicora&LL6&Chip&0&OC-TXH-014-A&C1OC14A&Q&$-$0.1272\\
&&125--500 \micron&0&OC-TXH-014-B&C1OC14B&Q&$-$0.0863\\
&&\textless125 \micron&0&OC-TXH-014-C&C1OC14C&Q&0.0344\\
  \hline
Cynthiana&L4&Chip&0&OC-TXH-015-A&C1OC15A&Q&$-$0.1183\\
&&125--500 \micron&0&OC-TXH-015-B&C1OC15B&Q&$-$0.0671\\
&&\textless125 \micron&0&OC-TXH-015-C&C1OC15C&Q&0.0682\\
&&\textless125 \micron\ pellet&0&OC-TXH-015-D&C1OC15D&Q&$-$0.0945\\
&&\textless125 \micron\ pellet&5&OC-TXH-015-D05&C1OC15D05&Sq&$-$0.0111\\
&&\textless125 \micron\ pellet&15&OC-TXH-015-D15&C1OC15D15&S&0.117\\
  \hline
Hedjaz&L3.7-L6&Chip&0&OC-TXH-016-A&C1OC16A&Q&$-$0.1057\\
&&Chip&20&OC-TXH-016-A20&C1OC16A20&Sq&$-$0.0752\\
&&Chip&40&OC-TXH-016-A40&C1OC16A40&S&$-$0.0202\\
&&125--500 \micron&0&OC-TXH-016-B&C1OC16B&Q&$-$0.1246\\
&&\textless125 \micron&0&OC-TXH-016-C&C1OC16C&Q&0.0413\\
  \hline
Soko-Banja &LL4&Chip&0&OC-TXH-017-A&C1OC17A&Q&$-$0.0794\\
&&125--500 \micron&0&OC-TXH-017-B&C1OC17B&O&$-$0.0460\\
&&\textless125 \micron&0&OC-TXH-017-C&C1OC17C&Q&0.0821\\
  \hline
Nulles&H6&Chip&0&OC-TXH-018-A&C1OC18A&Q&$-$0.1109\\
&H6&125--500 \micron&0&OC-TXH-018-B&C1OC18B&O&$-$0.1064\\
&&\textless125 \micron&0&OC-TXH-018-C&C1OC18C&Q&0.0197\\
&&\textless125 \micron\ pellet&0&OC-TXH-018-P&C1OC18P&Q&$-$0.1403\\
&&\textless125 \micron\ pellet&5&OC-TXH-018-P05&C1OC18P05&Q&$-$0.0655\\
&&\textless125 \micron\ pellet&15&OC-TXH-018-P15&C1OC18P15&Sq&0.0562\\
&&\textless125 \micron\ pellet&35&OC-TXH-018-P35&C1OC18P35&S&0.1616\\
&&\textless125 \micron\ pellet&55&OC-TXH-018-P55&C1OC18P55&Sw&0.2729\\
  \hline
Olmedilla de Alarcon &H5&Chip&0&OC-TXH-019-A&C1OC19A&Q&$-$0.1282\\
&&125--500 \micron&0&OC-TXH-019-B&C1OC19B&Q&$-$0.1413\\
&&\textless125 \micron&0&OC-TXH-019-C&C1OC19C&Q&0.0076\\
&&\textless125 \micron\ pellet&0&OC-TXH-019-P&C1OC19P&Q&$-$0.1998\\
\footnotemark[$\dagger$] &&\textless125 \micron\ pellet&5&OC-TXH-019-P05&C1OC19P05&Sq&$-$0.1445\\
\footnotemark[$\dagger$] &&\textless125 \micron\ pellet&15&OC-TXH-019-P15&C1OC19P15&S&$-$0.0818\\
  \hline
Dhajala\footnotemark[$*$]&H3.8&Chip&0&OC-TXH-020-A&C1OC20A&Sq&$-$0.0340\\
\footnotemark[$*$]&&125--500 \micron&0&OC-TXH-020-B&C1OC20B&Sq&$-$0.0161\\
\footnotemark[$*$]&&\textless125 \micron&0&OC-TXH-020-C&C1OC20C&Sq&0.0916\\
  \hline
Burnwell\footnotemark[$*$]&H4-an&Chip&0&OC-TXH-021-A&C1OC21A&Q&$-$0.1247\\
\footnotemark[$*$]&&Chip&20&OC-TXH-021-A20&C1OC21A20&K&$-$0.0174\\
\footnotemark[$*$]&&Chip&40&OC-TXH-021-A40&C1OC21A40&Sq&$-$0.0611\\
\footnotemark[$*$]&&125--500 \micron&0&OC-TXH-021-B&C1OC21B&Q&$-$0.1498\\
\footnotemark[$*$]&&\textless125 \micron&0&OC-TXH-021-C&C1OC21C&Sq&$-$0.0071\\
\end{longtable}
\addtocounter{table}{-1}

\begin{figure*}
  \begin{center}
    \FigureFile(56mm,56mm){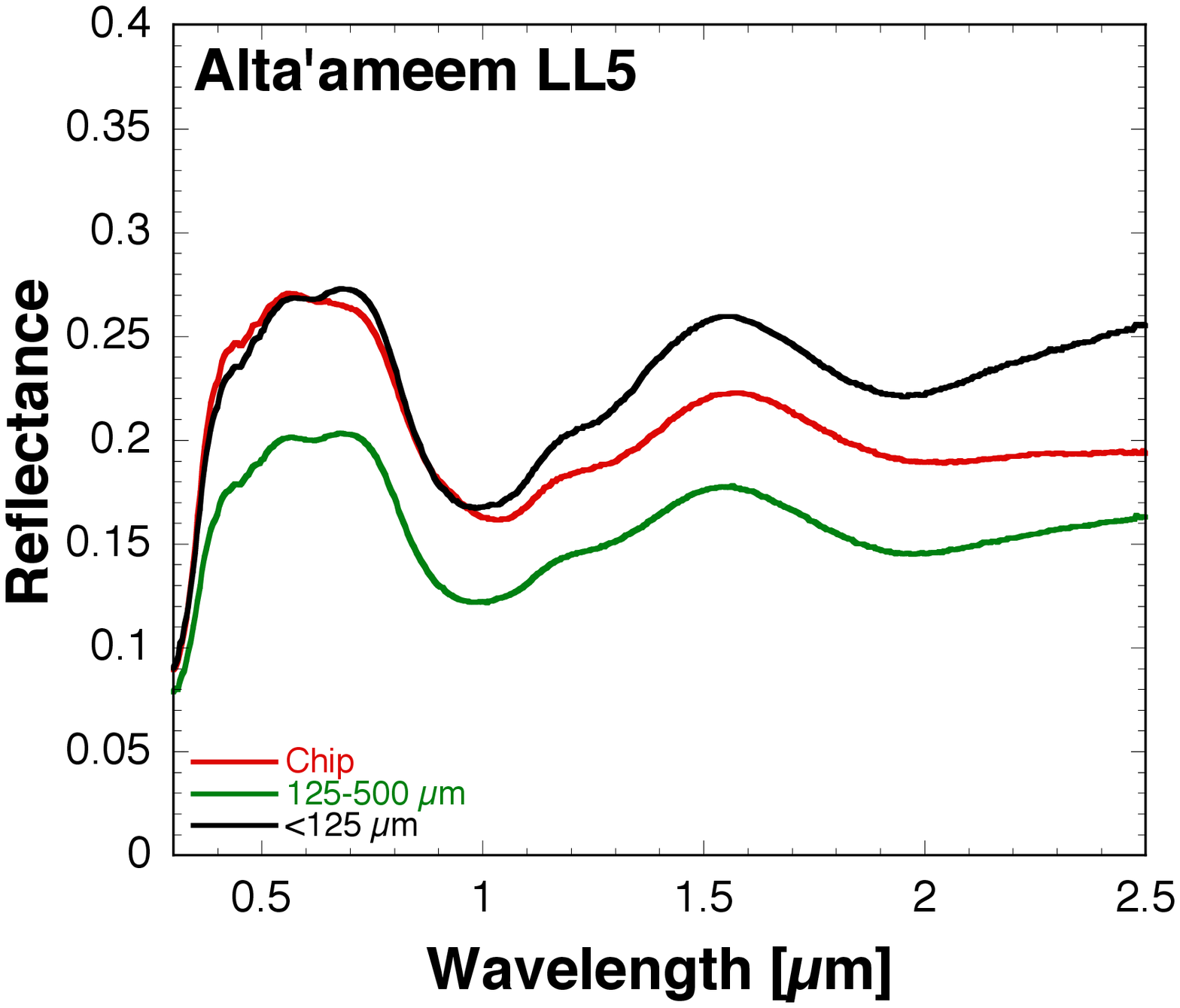}
    \FigureFile(56mm,56mm){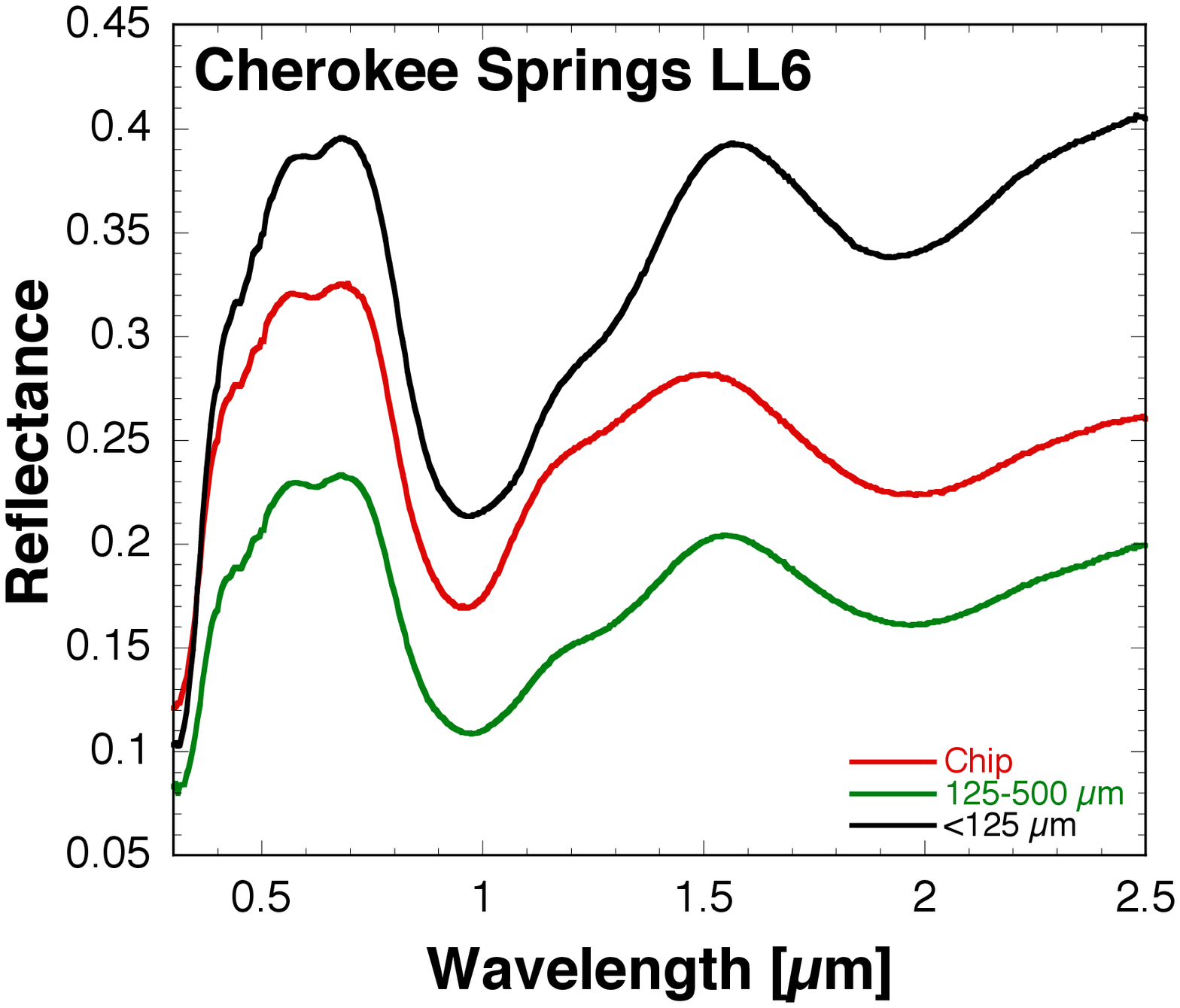}
    \FigureFile(56mm,56mm){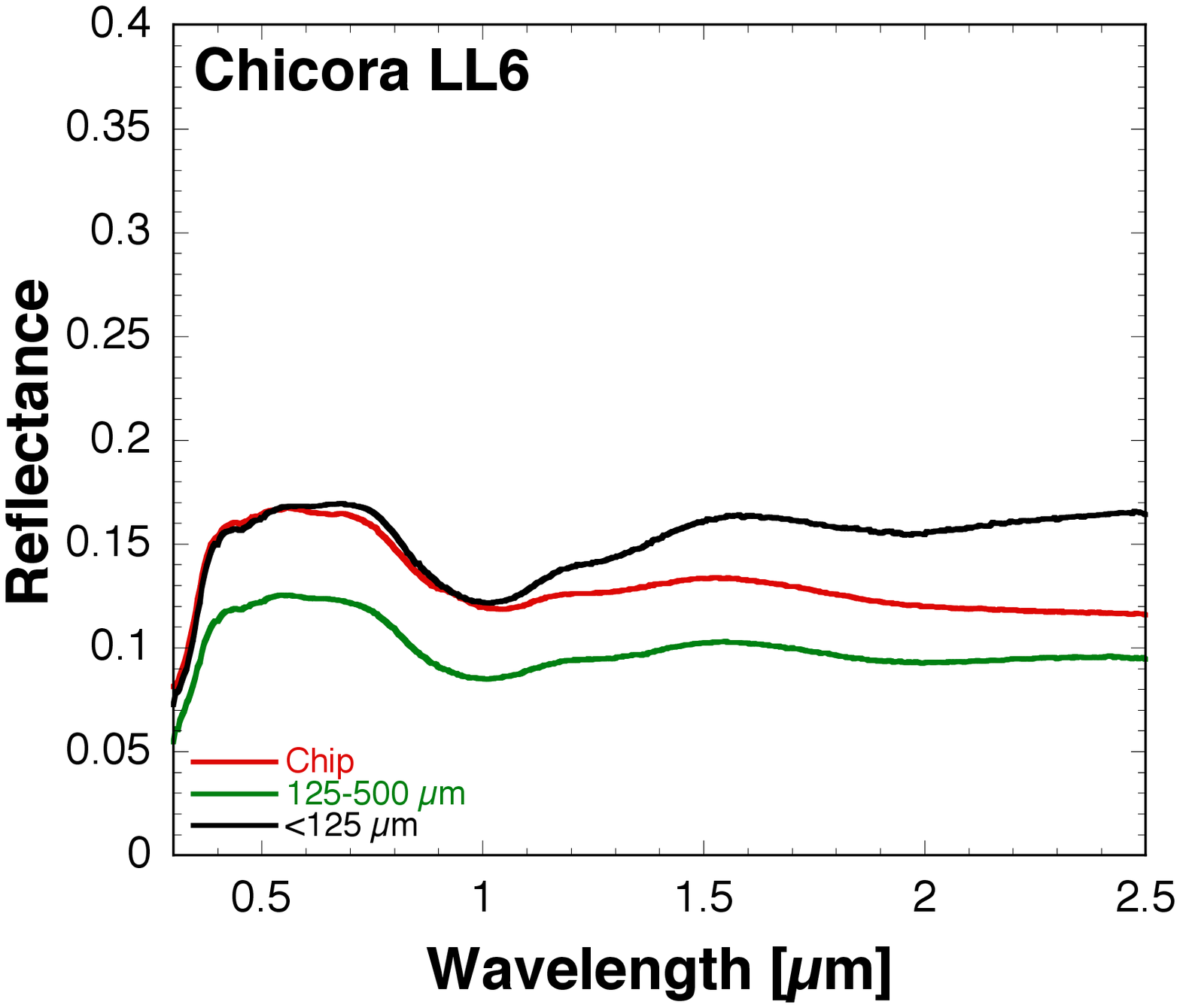}
    \FigureFile(56mm,56mm){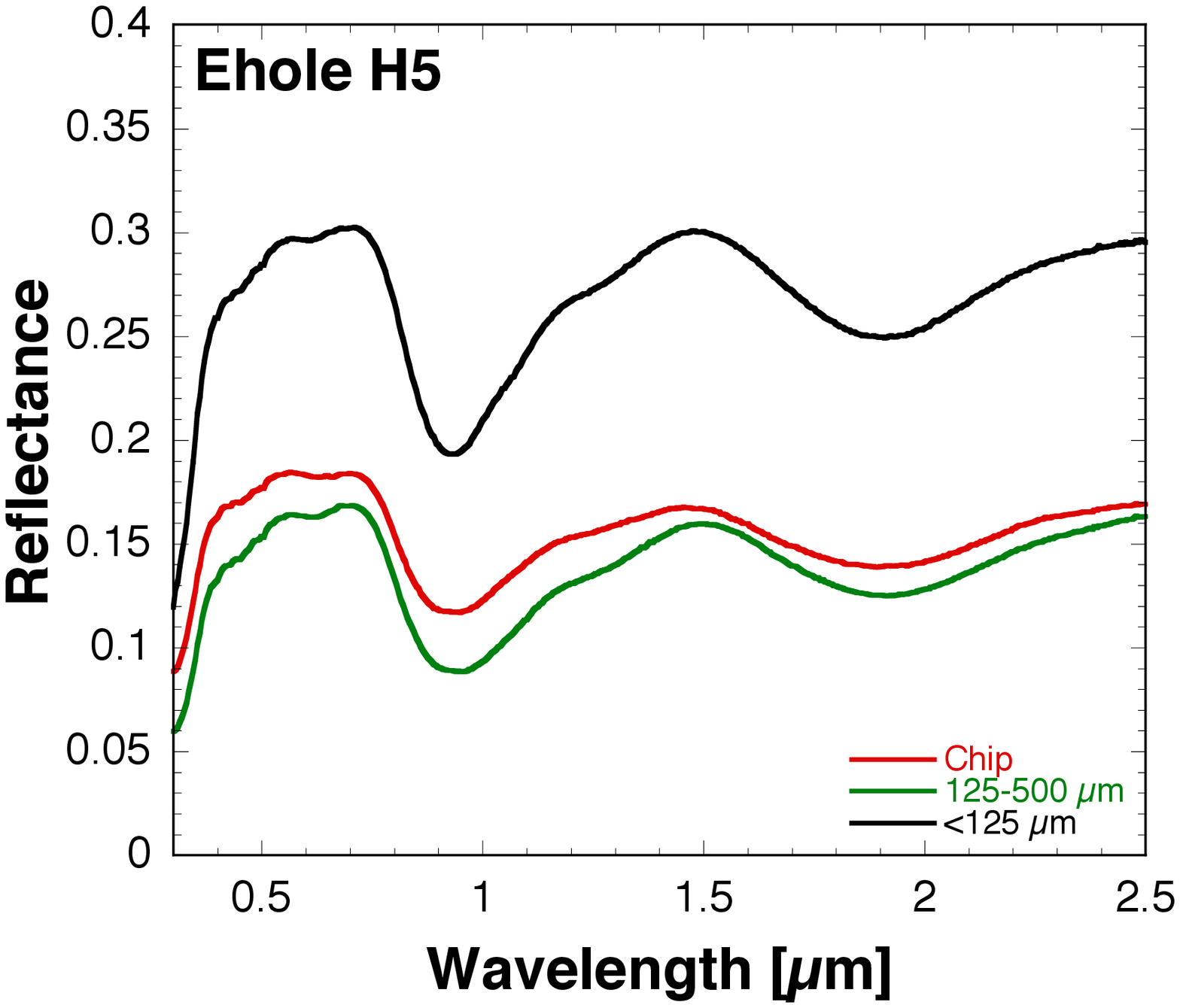}
    \FigureFile(56mm,56mm){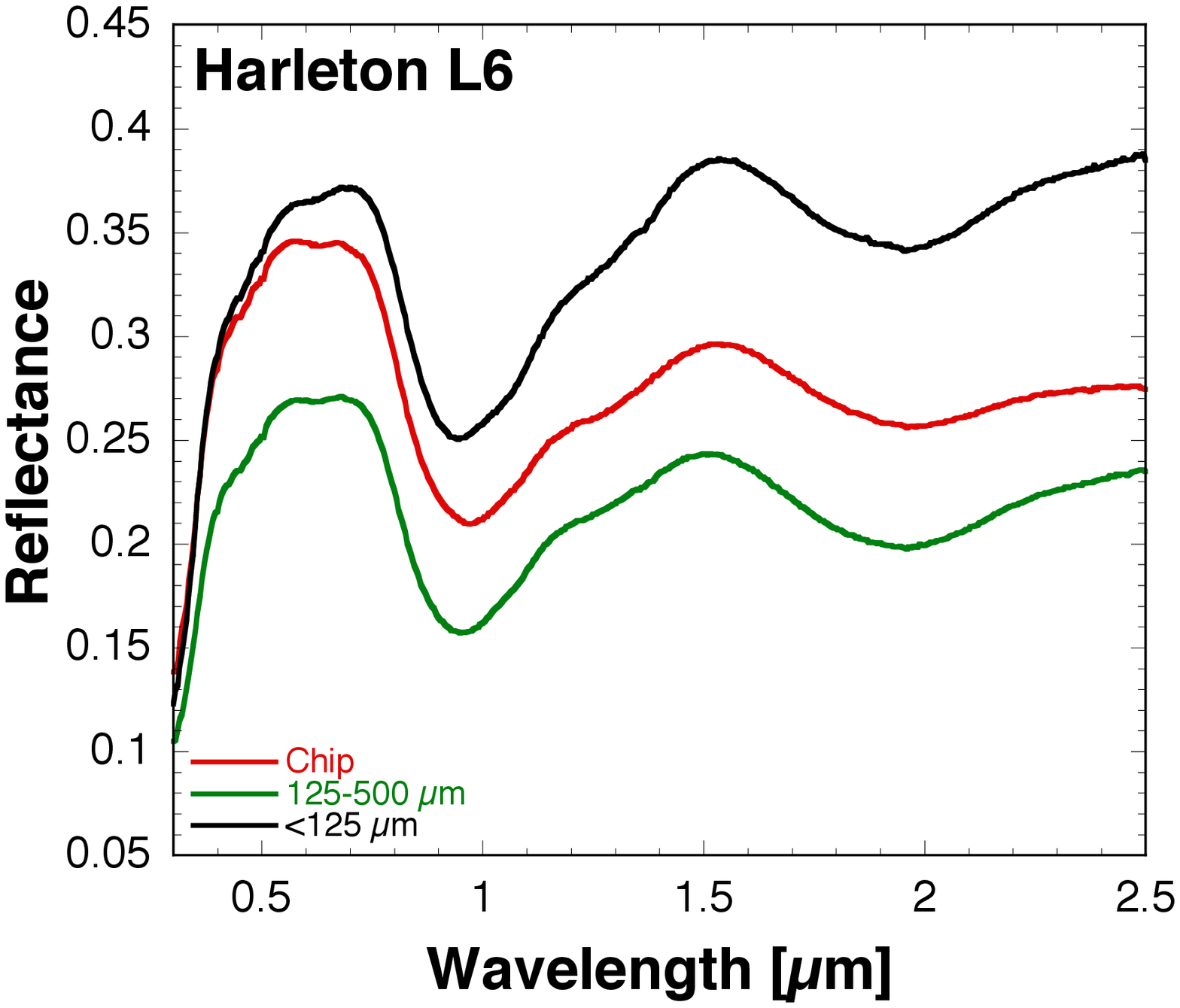}
    \FigureFile(56mm,56mm){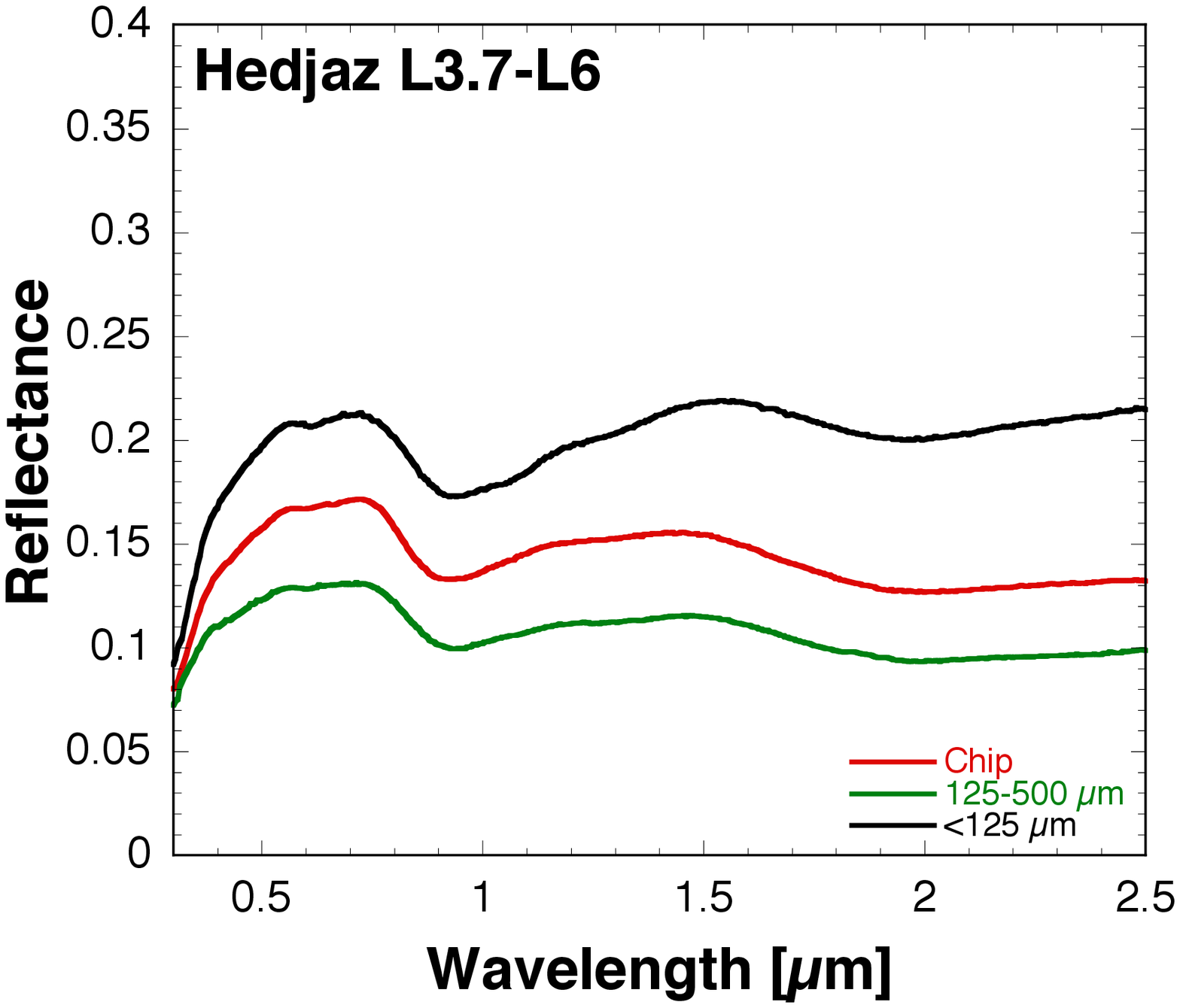}
    \FigureFile(56mm,56mm){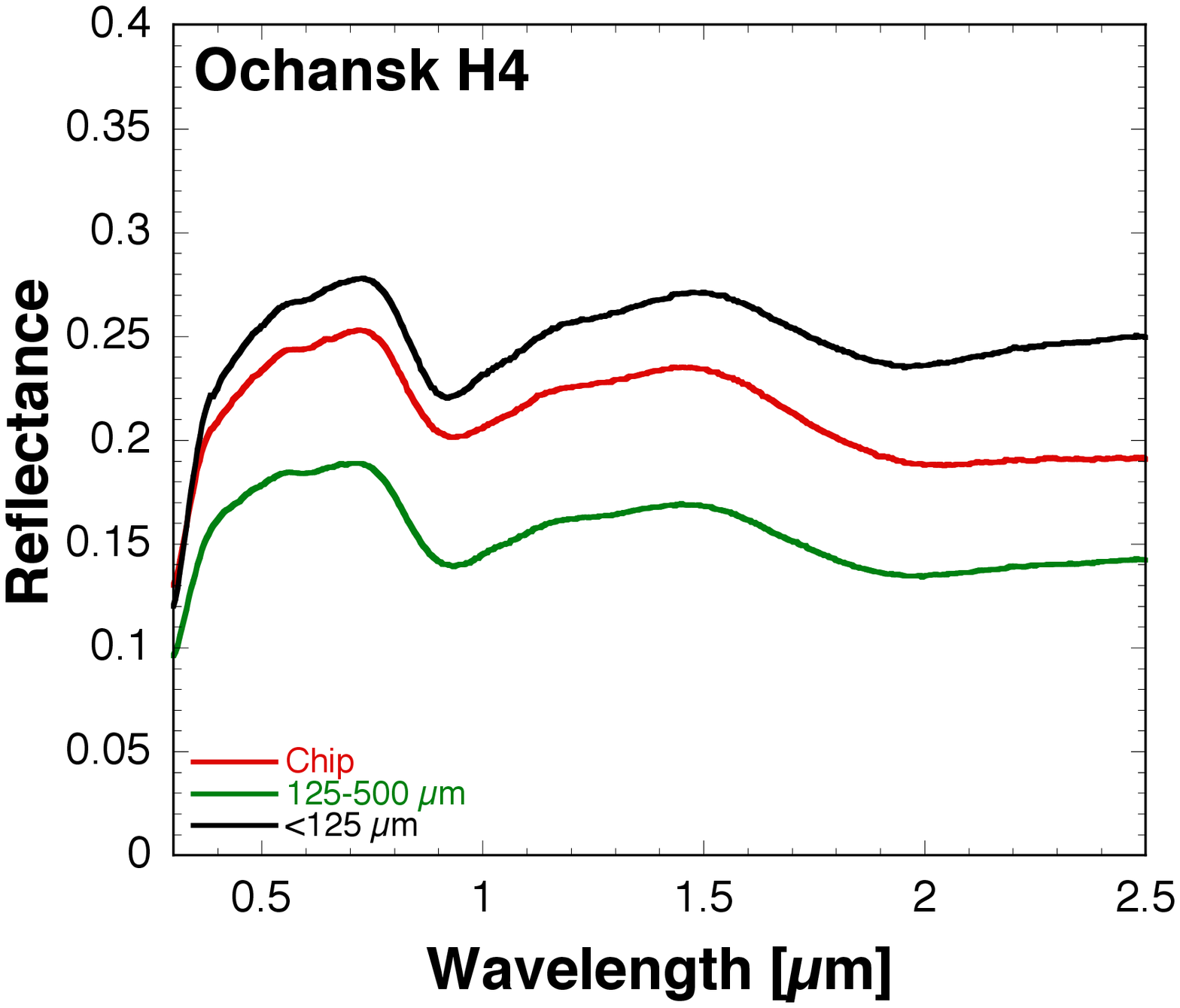}
    \FigureFile(56mm,56mm){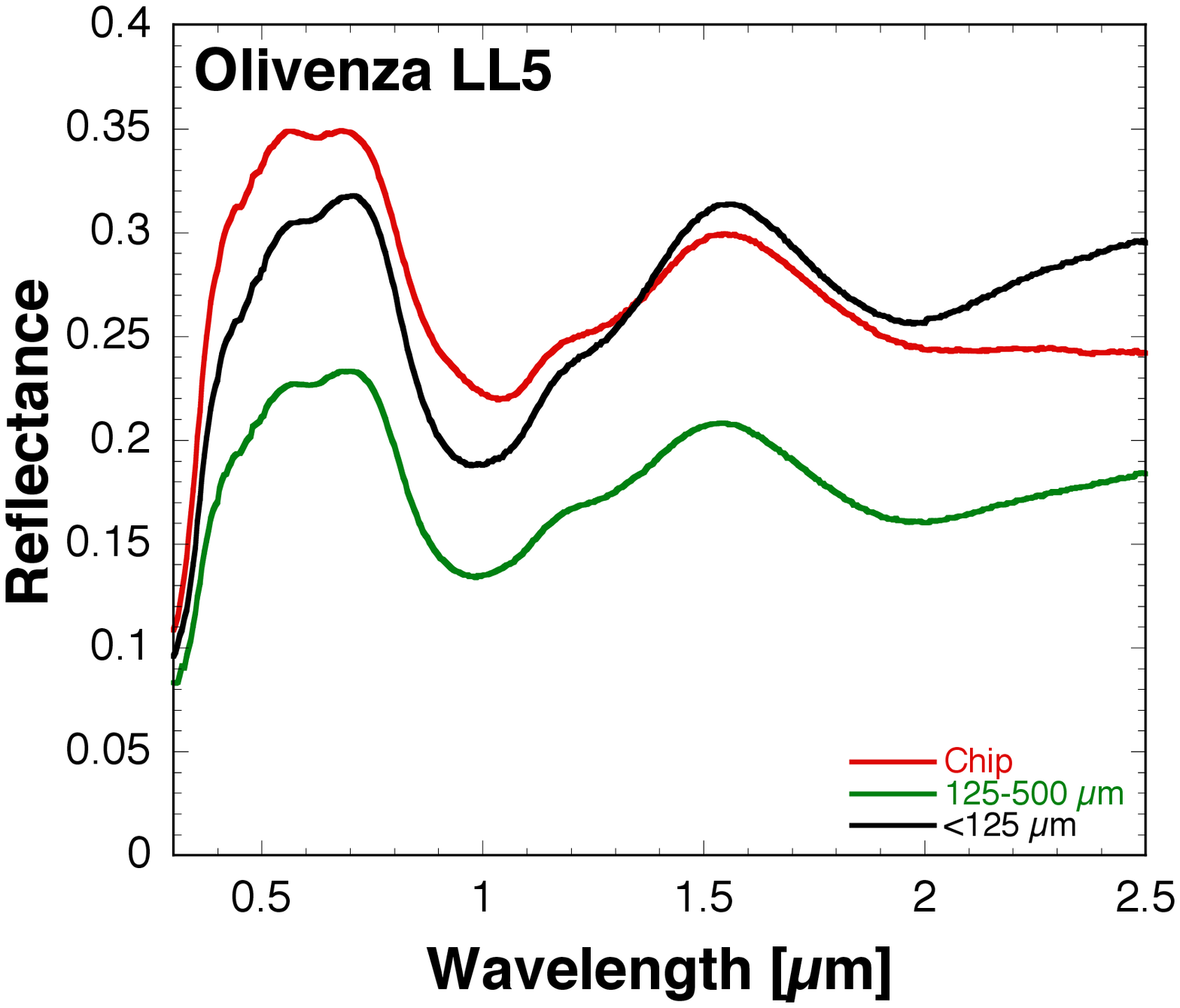}
    \FigureFile(56mm,56mm){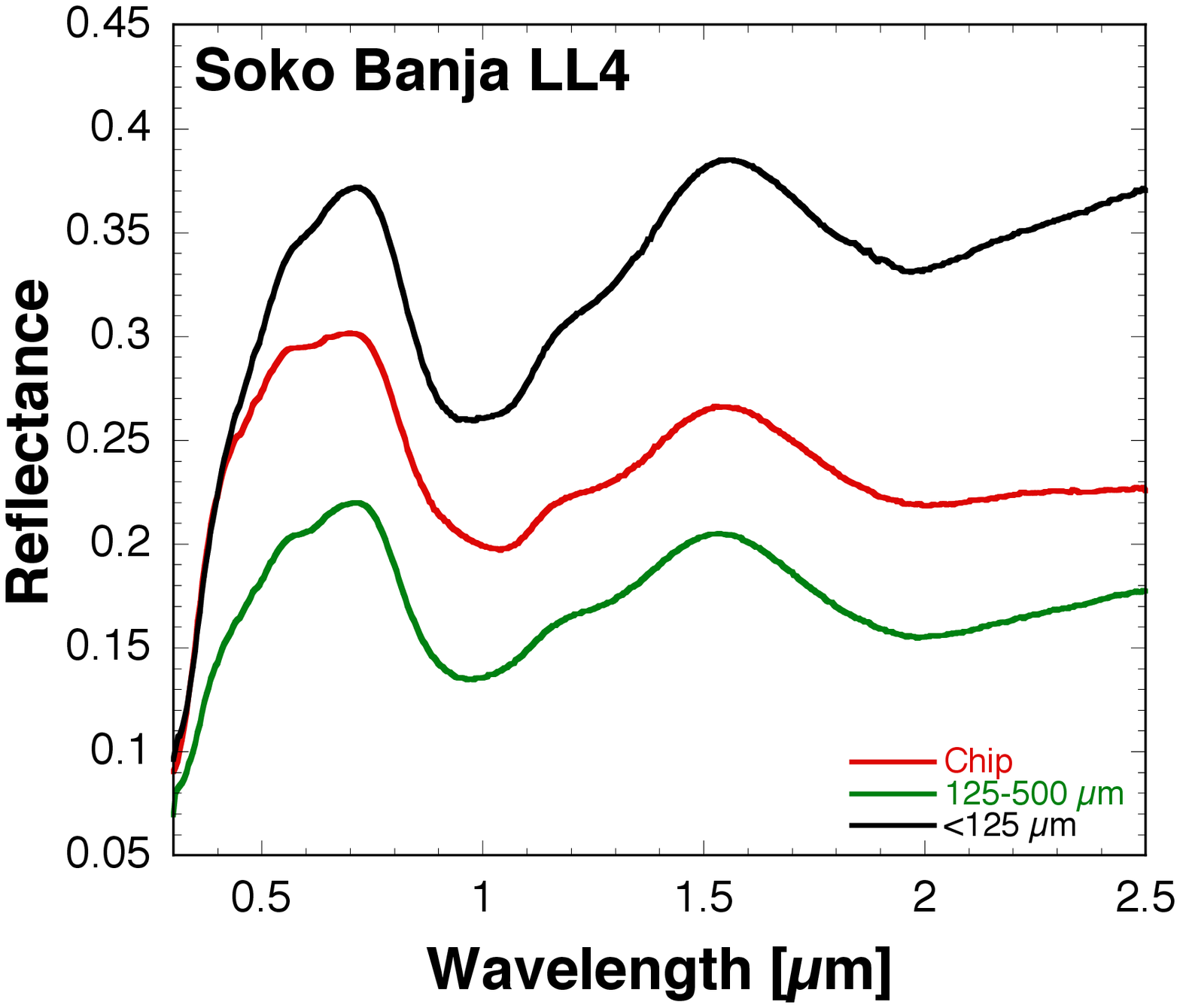}
  \end{center}
  \caption{Reflectance spectra of ordinary chondrites measured at an incident angle of 30 degrees and reflection angle of 0 degrees, demonstrating the effect of the particle size on the spectra of ordinary chondrites; with the increase in particle size, the appearance of bluish spectral slopes increases.
}
\label{fig:NL}
\end{figure*}

\begin{figure*}
  \begin{center}
    \FigureFile(56mm,56mm){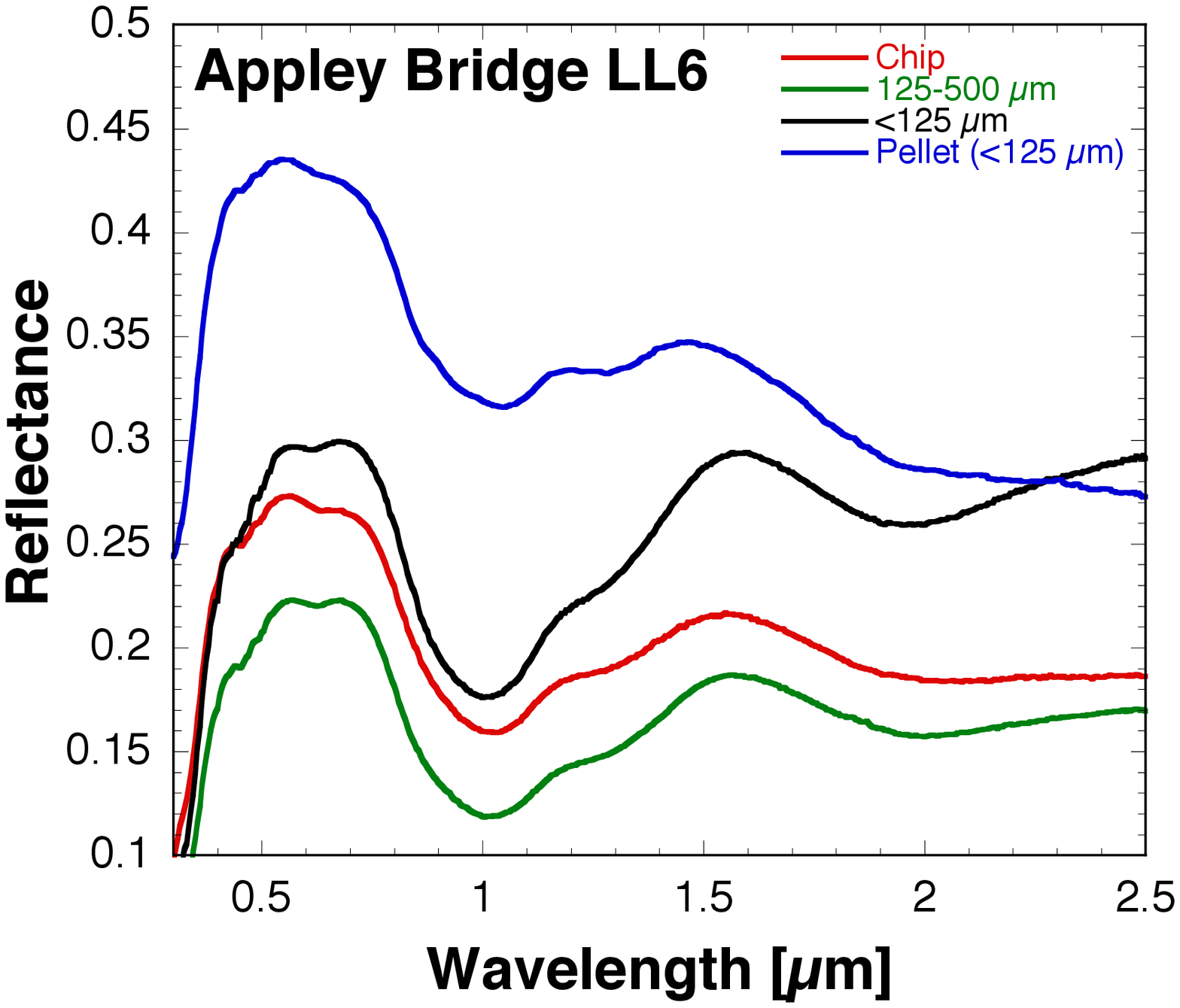}
    \FigureFile(56mm,56mm){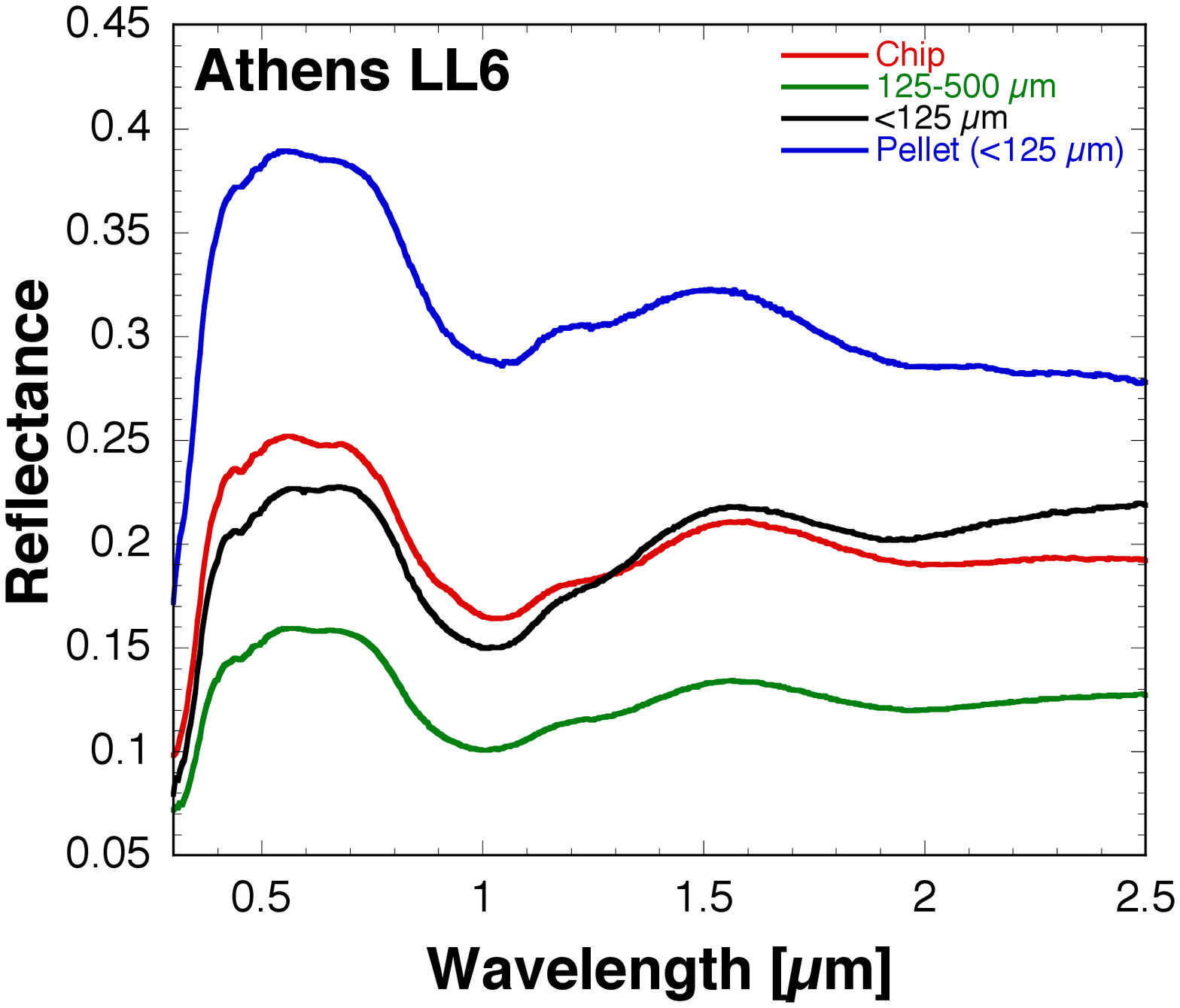}
    \FigureFile(56mm,56mm){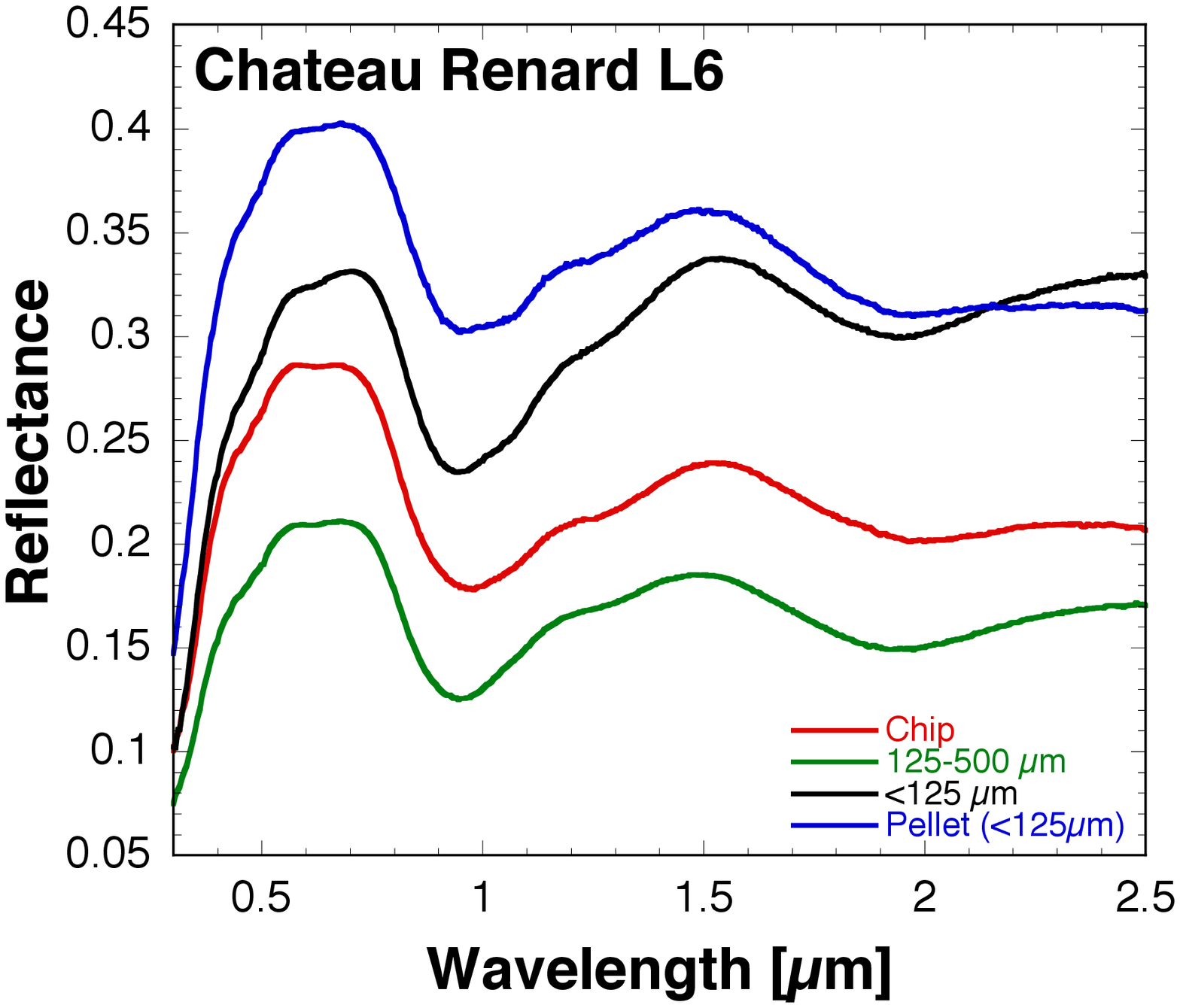}
    \FigureFile(56mm,56mm){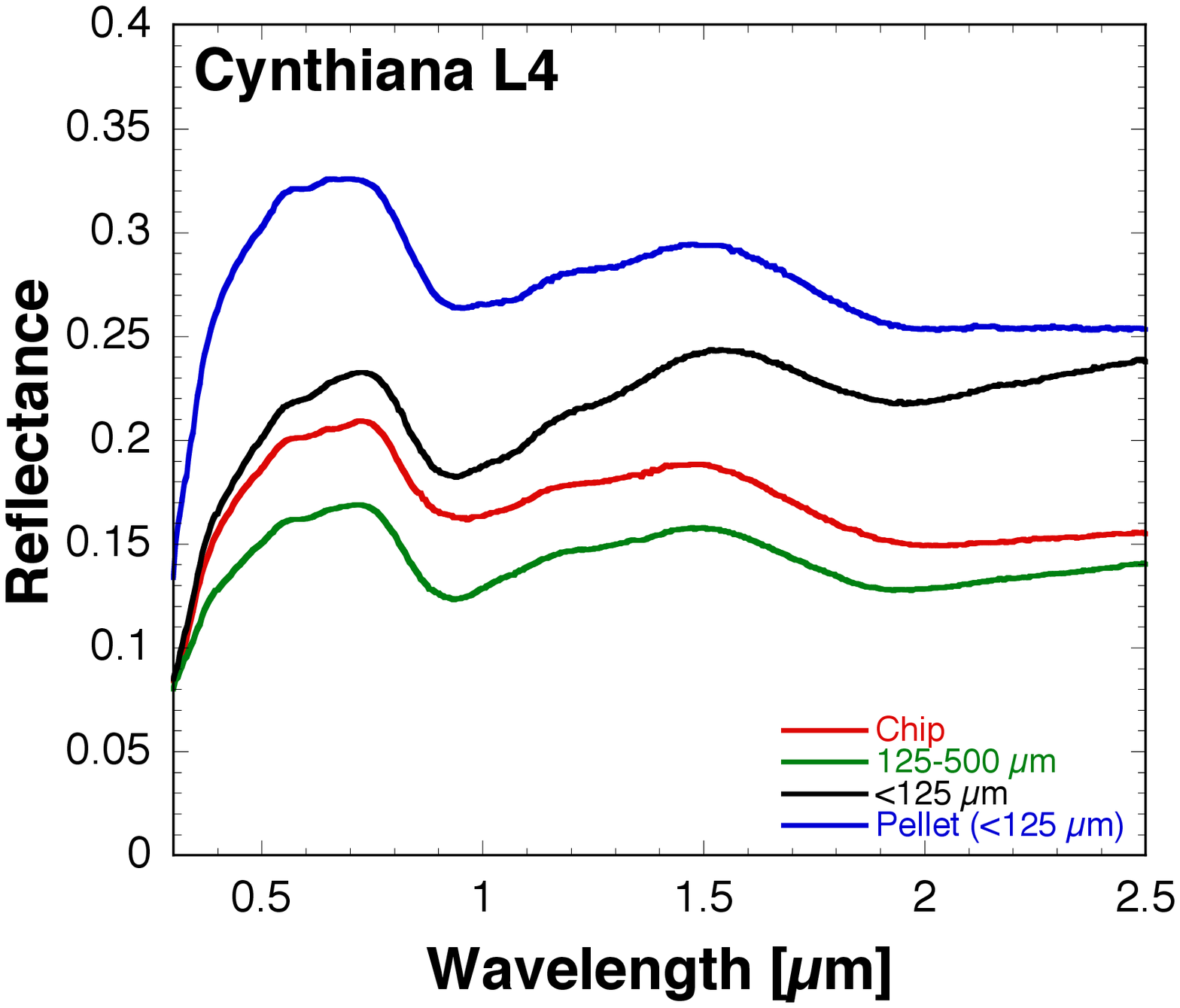}
    \FigureFile(56mm,56mm){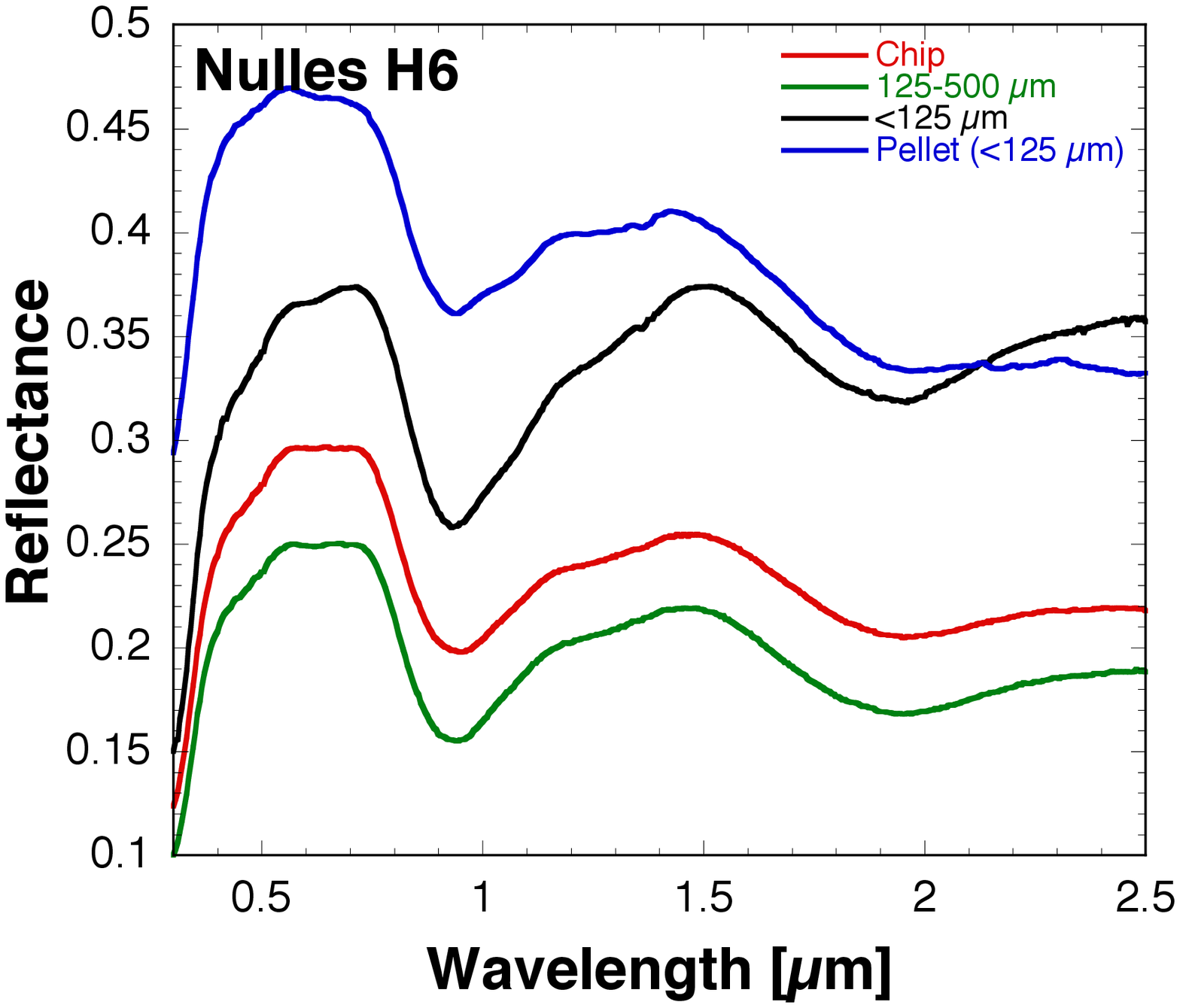}
    \FigureFile(56mm,56mm){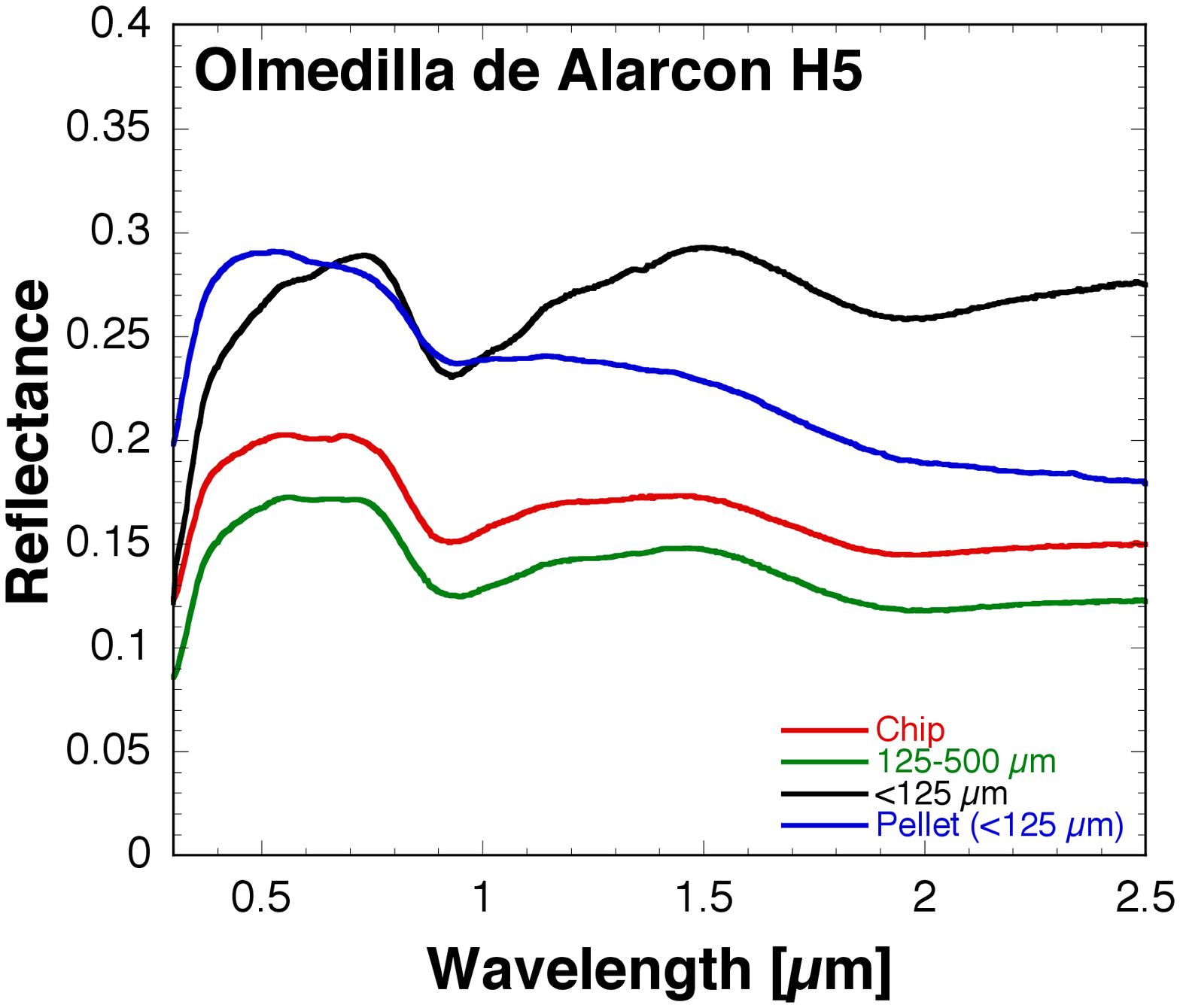}
  \end{center}
  \caption{Reflectance spectra of ordinary chondrites measured at incident ange: 30 degrees and reflection angle: 0 degrees.
This figure shows the spectra of the pellet samples, which is bluer than the other spectra.
}
\label{fig:NLP}
\end{figure*}

\begin{figure*}
  \begin{center}
    \FigureFile(56mm,56mm){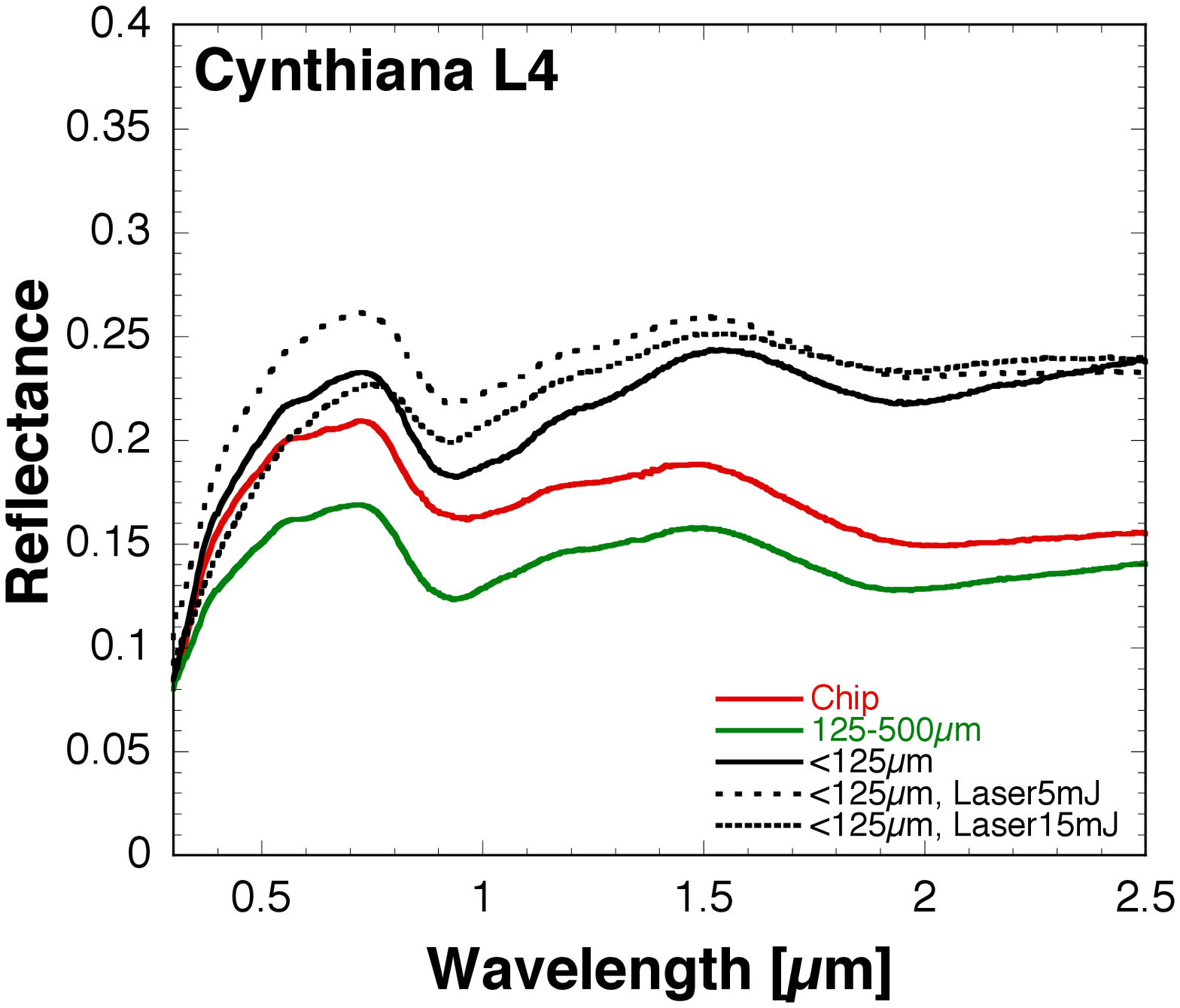}
    \FigureFile(56mm,56mm){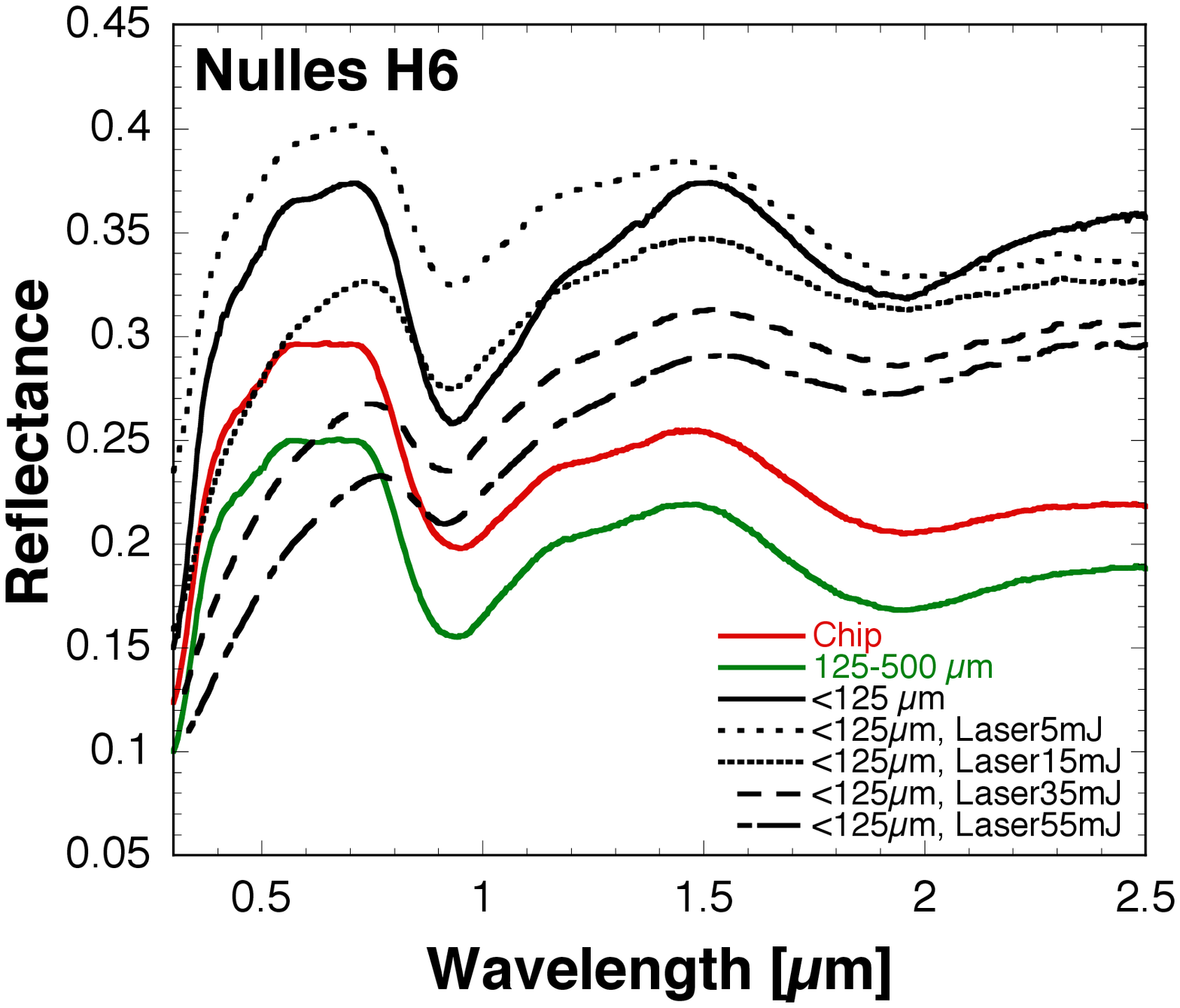}
    \FigureFile(56mm,56mm){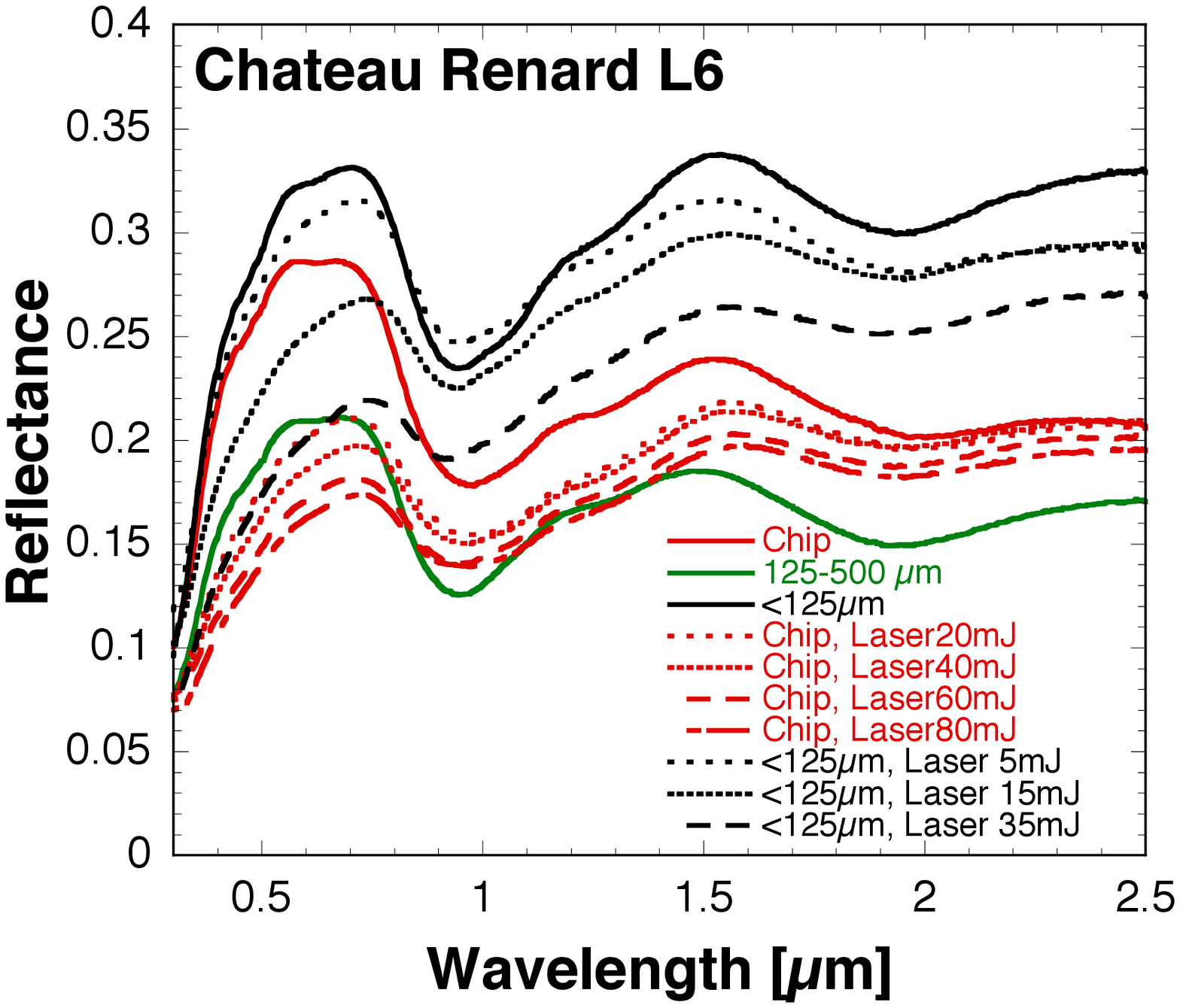}
    \FigureFile(56mm,56mm){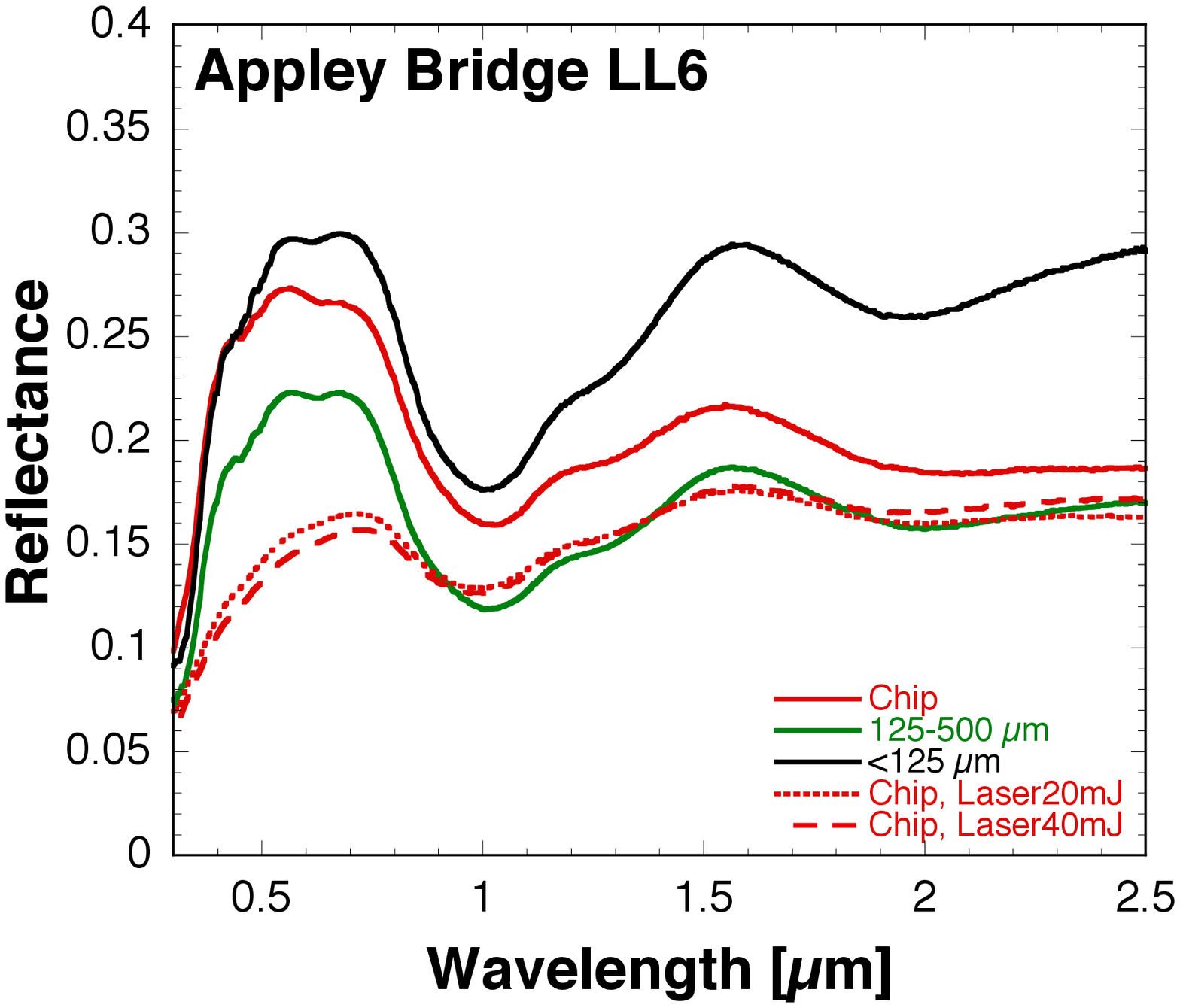}
    \FigureFile(56mm,56mm){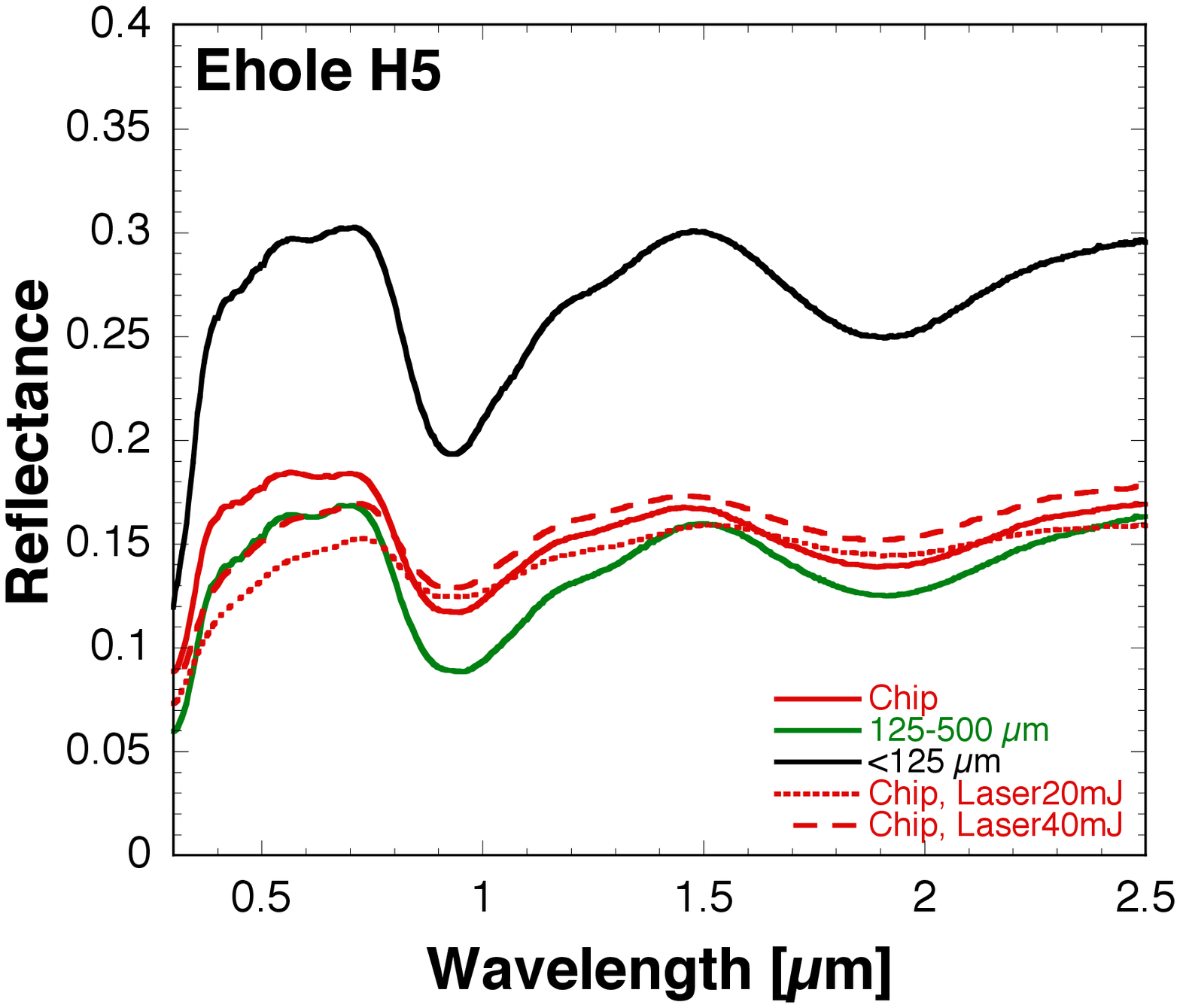}
    \FigureFile(56mm,56mm){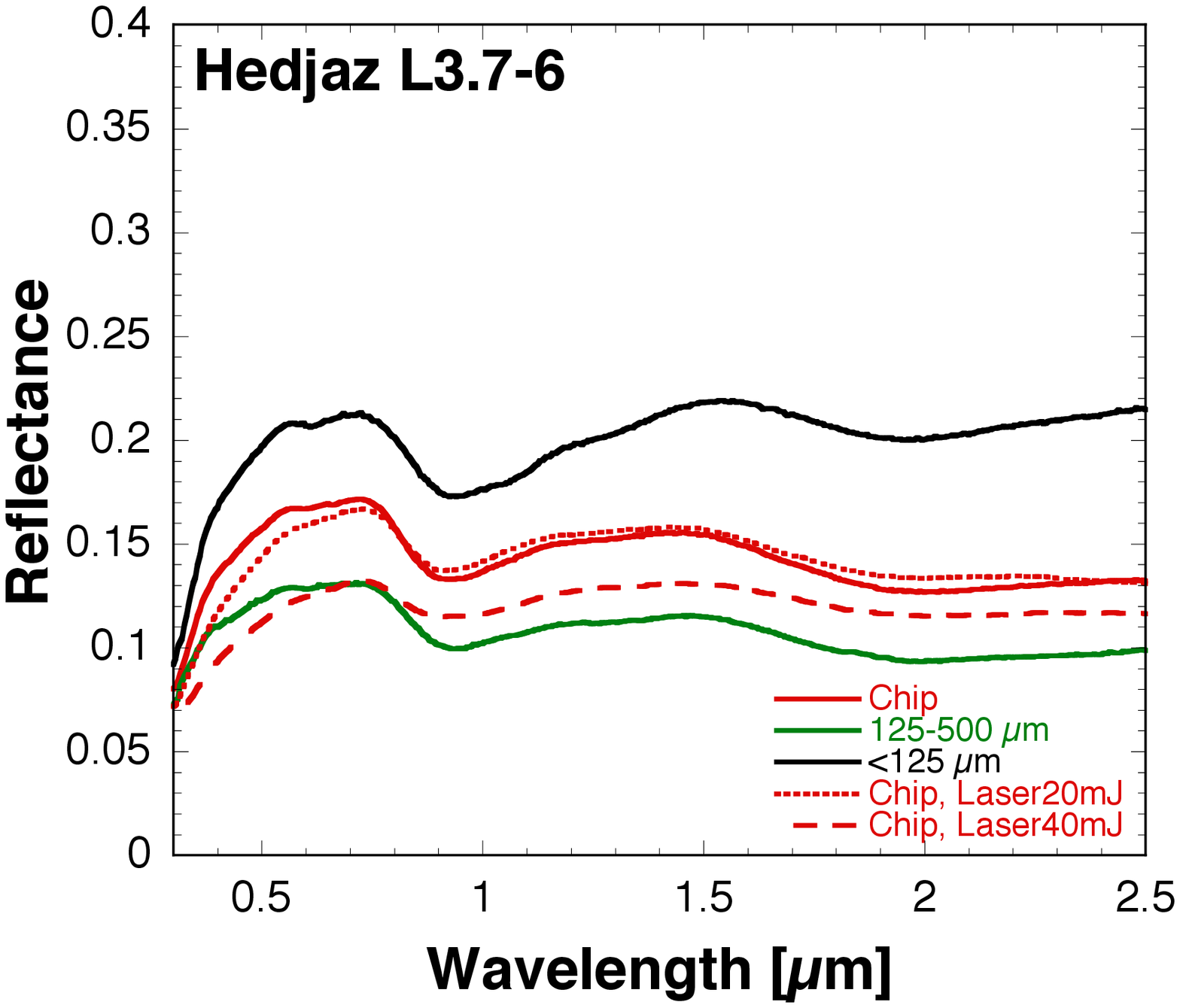}
  \end{center}
  \caption{Reflectance spectra with the space weathering process for ordinary chondrites measured at an incident angle of 30 degrees and reflection angle of 0 degrees, demonstrating the effect of the particle size and space weathering on the spectra of ordinary chondrites.
With the increase in particle size and degree of space weathering, the appearance of bluish and reddish spectral slopes increase, respectively.
In particular, the chip sample retains the Q-type spectrum, even with the progress of space weathering.
}
\label{fig:W}
\end{figure*}

Initially, the effect of the particle size on the spectra was examined for the meteorite samples before space weathering.
Increasing the particle size of ordinary chondrites resulted in the appearance of more blue-sloped spectra (figures \ref{fig:NL} and \ref{fig:NLP}).
The spectra of particles less than 125 \micron\ was particularly red; conversely, the spectra of particles between 125--500 \micron\ and the chip samples included blue slopes.
The mean values of the slope of the Bus-DeMeo taxonomy for the chip, 125--500 \micron, and \textless125 \micron\ samples were $-$0.096 $\pm$ 0.030, $-$0.076 $\pm$ 0.041, and 0.039 $\pm$ 0.033, respectively.
The mean slope of particles between 125--500 \micron\ was consistent with that of the chips, within the standard deviation.
This indicates that the spectral slope is influenced by the presence or absence of particles of approximately 100 \micron\ or less.

Pressure was applied to pellets that had the fineness of powder, after which the surface was scraped and made diffusive; however, it exhibited the same surface reflection characteristics as that of a uniformly homogenized chip surface.
Hence, the slope of the pellet-sample spectrum was in the negative direction, and the brightness increased (see figure \ref{fig:NLP}).
Therefore, it was difficult to adapt the reflectance of the pellet sample to that of the powder.
However, pellets without obviously bluish chips were applied as powder for space weathering in this study.
As a result, three space weathered pellet samples were available.

Figure \ref{fig:W} shows that the increase in the degree of space weathering toward that of ordinary chondrites leads to the appearance of more red-sloped spectra. 
The spectra of weathered ordinary chondrites composed of particles below 125 \micron\ had reddish slope, but the slope of the weathered chip spectra tended to be flat.
Based on the Bus-DeMeo taxonomy \citep{DeMeo2009}, most of the weathered ordinary chondrites within 125 \micron\ were classified as S-complex, whereas most of the chip samples were classified as Q-type (see table \ref{tab:M}).

\section{Discussion}
The results of the numerical orbit propagation in this study indicate that the surface refreshening of only approximately 50 percents of Q-type asteroids can be explained by planetary encounter even with the best estimate (see section 2).

The spectral measurements indicate that weathered ordinary chondrites with chip surfaces can be judged as Q-type asteroids (See section 3).
Moreover, it has also been demonstrated that the spectral slope is similar for the chip and 125--500 \micron\ samples.
These results indicate that ordinary chondritic asteroids without less than 100 \micron\ particles on the surface can be judged as Q-type under space weathering.

However, a mechanism to release less than 100 \micron\ particles from the asteroid surface layer is required.
There are several methods for this mechanism, including centrifugal force, electrostatic acceleration and/or solar radiation pressure.

Below is an approximate estimate of the sizes of the particles that can escape from each Q-type asteroid due to centrifugal force.
Assuming that the asteroid is a spherical object and is noncohesive, it is considered that the centrifugal force and gravity on the surface of the object are balanced.
The gravity and centrifugal force on the asteroid surface are balanced, when the bulk density of the asteroid is 2000 kg $\mathrm{m^{-3}}$ from upper-limit density value of asteroid 25143 Itokawa \citep{Fujiwara2006}, and the rotational period is \textit{P} = 2.3 hr.
Of the 101 Q-type asteroids with known rotational periods (\authorcite{Benishek2016} \yearcite{Benishek2016}, \yearcite{Benishek2018}; \cite{Birtwhistle2009}; \cite{Durech2018}; \cite{Galad2005}; \cite{Galad2010}; \cite{Hasegawa2018}; \cite{Hergenrother2009}; \authorcite{Hicks2013} \yearcite{Hicks1998}, \yearcite{Hicks2012}, \yearcite{Hicks2013}; \cite{Higgins2006}; \cite{Hills2014}; \cite{Koehn2014}; \cite{Larsen2014}; \cite{Mottola1995}; \cite{Oberst2001}; \cite{Oey2014}; \cite{Oey2017}; \cite{Pilcher2012}; \cite{Polishook2012}; \authorcite{Pravec1997} \yearcite{Pravec1997}, \yearcite{Pravec2005}, \yearcite{Pravec2006}; \cite{Pravec1996}; \cite{Rozitis2013}; \cite{Skiff2012}; \cite{Stephens2014}; \cite{Thirouin2016}; \cite{Vaduvescu2017}; \authorcite{Warner2012a} \yearcite{Warner2012a}, \yearcite{Warner2012b}, \yearcite{Warner2014a}, \yearcite{Warner2014b}, \yearcite{Warner2014c}, \yearcite{Warner2015a}, \yearcite{Warner2015b}, \yearcite{Warner2015c}, \yearcite{Warner2015d}, \yearcite{Warner2016a}, \yearcite{Warner2016b}, \yearcite{Warner2017a}, \yearcite{Warner2017b}, \yearcite{Warner2018a}, \yearcite{Warner2018b}; \cite{Warner2014d}; \cite{Warner2013}; \authorcite{Warner2019a} \yearcite{Warner2019a}, \yearcite{Warner2019b}; \cite{Warner2015e}; \cite{Waszczak2015}; \cite{Wisniewski1991}; \cite{Wisniewski1997}; \cite{Ye2009}; Goldstone Radar Observations Planning website\footnotemark[3]; Observatoire de Geneve web site\footnotemark[4]; the Ondrejov Asteroid Photometry Project website\footnotemark[5]; Posting on CALL web site at the Collaborative Asteroid Lightcurve Link\footnotemark[6]), only six have rotation cycles less than 2.3 hr.
Therefore, it is also difficult to explain the surface refreshening of all the Q-type asteroids through the centrifugal force.

\footnotetext[3]{https://echo.jpl.nasa.gov/asteroids/2002NV16/2002NV16\%5Fplanning.html}
\footnotetext[4]{http://obswww.unige.ch/\%7Ebehrend/page\%5Fcou.html}
\footnotetext[5]{http://www.asu.cas.cz/\%7Eppravec/neo.htm}
\footnotetext[6]{http://www.minorplanet.info/call.html}

\begin{figure*}
  \begin{center}
    \FigureFile(80mm,80mm){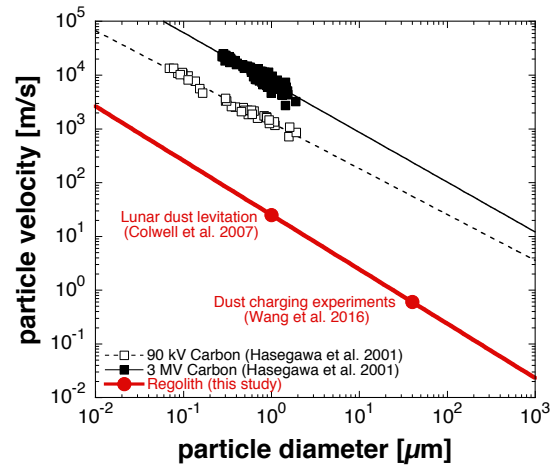}
  \end{center}
  \caption{The relationship between the particle size and particle velocity.
Conductor (Carbon) data are obtained from \citet{Hasegawa2001}.
The relationship between the particle size and velocity for regolith particles is similar to that of conductor particles.
}
\label{fig:C}
\end{figure*}

Below is an approximate estimate of the particle sizes of the escaped particles from the surface through electrostatic acceleration.
\citet{Colwell2007} showed that micron-sized lunar dust regolith reach heights of a few to hundreds of meters due to charge.
The micron-sized particle initial velocity is estimated as 25 m $\mathrm{s^{-1}}$ , based on this description.
\citet{Wang2016} experimentally found that 40-\micron\ particles accelerated to 0.6 m $\mathrm{s^{-1}}$.
By extrapolating between the lunar observations and experimental results, it can be estimated that 100-\micron\ particles are accelerated to a velocity of approximatley 0.2 m $\mathrm{s^{-1}}$ (see figure \ref{fig:C}).
If the escape velocity is lesser than this velocity, it is possible for the particles to escape from the surface due to electrostatic force.
Based on this estimation, from 64 small objects less than approximately 0.3 km in diameter, particles lesser than 100 \micron\ can escape from the surface.
\citet{Graves2019} indicated that the distribution of the spectral slopes of S-complex and Q-type asteroids decreases with the decrease in size.
This results also validates the escape mechanism due to electrostatic force.

Below is an approximate estimate of the particle sizes of the escaped particle from each Q-type asteroid due to solar radiation pressure.
For this purpose, Equation 19 from \citet{Burns1979} was used, which stated that particles escape from the surface when the solar radiation pressure exceeds the gravitational attraction.
It is assumed that particle escape on the asteroidal surface occurs at the perihelion.
%
%
The Q-type asteroid diameter is calculated from the absolute magnitude and mean value of the geometric albedo 0.27 from \citet{Hasegawa2017}.
The asteroidal mass is determined in combination with the diameter and asteroid density, which is considered to be 2000 kg $\mathrm{m^{-3}}$ based on the upper-limit density value of Itokawa \citep{Fujiwara2006}.
For the particle density, the average value of the bulk density for ordinary chondrite is used, which is 3400 kg $\mathrm{m^{-3}}$ \citep{Consolmagno2008}.
For the particle triaxiality ratio, the typical triaxiality ratio 1:$\surd$2:2 of Itokawa particles is used \citep{Michikami2018}.
It is assumed that particles are blown at the maximum cross-section.
As per the estimation, it is possible that less than 100-micron particles can escape from the surfaces of 31 small-sized Q-type asteroids that are less than approximately 0.5 km in diameter with a perihelion of less than 0.4 au.
\citet{Binzel2019} and \citet{Graves2019} demonstrated that the distribution of the spectral slopes of S-complex and Q-type asteroids decrease with the decrease in the perihelion and size.
This result validates the escape mechanism due to solar radiation pressure.

Due to solar radiation pressure or electrostatic force, less than 100-\micron\ particles can escape from the surfaces of less than 0.5 km-sized asteroids with a perihelion close to the sun, or from less than 0.3 km-sized asteroids.
However, 110 Q-type asteroids cannot be explained using the centrifugal force, solar radiation pressure, or electrostatic acceleration mechanisms alone. 
One possibility is the uneven distribution of particles on the asteroid surface.
On 25143 Itokawa, fine particles were observed in low-gravity areas \citep{Fujiwara2006}.
\citet{Tardivel2018} simulated the localization of particles on the asteroid surface by the effect of centrifugal force.
Therefore, small particles may be collected at low-potential local areas, and particles smaller than 100 \micron\ may be eliminated from most areas by centrifugal force in combination with solar radiation pressure and/or electrostatic force.

However, these mechanisms cannot explain the release of small particles on asteroids with large perihelion such as the Mars crossers with rotational period of more than 2.3 hr.
This may imply the need for other mechanisms for the escape of particles lesser than 100 \micron\ from asteroid surfaces, or additional concepts such as particle convection (\authorcite{Yamada2016} \yearcite{Yamada2016}, \yearcite{Yamada2017}) to form the spectra of Q-type asteroids.

A few km Q-type asteroids were found also among young asteroids involved in asteroid pairs in the main belt (\cite{Polishook2014}; \cite{Pravec2019}). 
Solar radiation pressure and electrostatic force do not affect small particle exclusion on these km-sized non near-Earth asteroids.
However, since these asteroids were temporarily binary systems (before the secondaries were ejected; see \cite{Pravec2019}, and references therein), three mechanisms can be considered. 
One possibility is that radiation pressure and electrostatic force are expected to work effectively to particle release on the binary asteroids because centrifugal force and gravity are expected to be nearly balanced on the binary formed by the rotation fission.
Another possibility is that large particles may float to the surface by the Brazil nut effect (e.g., \cite{Perera2016}) excited by tiny tidal force on the binary system.
The other possibility is that surface circulation may occur by the tidal force, and a non-weathered surface may be replaced in several thousand years. 
However, these mechanism may be unique to the binary systems, and do not affect for non-binary asteroids.

In the space weathering experiment, it changed from O-type to Q-type, but not vice-versa.
This result implies that the O-type asteroid has a fresh surface.
However, asteroids classified as O-type in near-Earth asteroids are asteroids using the Bus taxonomy \citep{Bus2002} in only the visible wavelength region (\cite{Perna2018}; \cite{Binzel2019}; \cite{Popescu2019}).
Near-Earth asteroids classified as O-type in the classification of \citet{DeMeo2009} including near-infrared data, have not been found at present.

Focusing on individual asteroids, the asteroid 68063 2000 $\mathrm{YJ_{66}}$ and 143624 2003 $\mathrm{HM_{16}}$ have obvious blue-sloped spectra in the visible to near-infrared wavelength regions (\cite{Binzel2019}; \cite{Popescu2019}).
These objects may have fresh surface covered with larger than 100-\micron\ particles.

\section{Conclusions}
Itokawa particles not only offered a solution to the S-complex asteroid problem, but also created a new ambiguity that the space weathering time scale was three orders of magnitude lesser than the previously considered scale. 
Even if it estimate optimistically as much as possible, the orbital evolution calculation in this study revealed that the mechanism for refreshing the asteroid surface by close encounters with planets can be applied maximally to only half the currently known Q-type asteroids, approximately. 
The results of laboratory experiments on space weathering demonstrated that ordinary chondrites without less than 100-\micron\ particles can be established as Q-types, even with space weathering. 
Solar radiation pressure and electrostatic force can be considered as mechanisms for releasing particles lesser than 100 \micron\, and it was determined that they can be applied to Q-type asteroids less than 0.5 km in diameter with very small perihelion or Q-type asteroids less than 0.3 km in diameter.

\bigskip
\begin{ack}
We would like to thank Dr. Petr Pravec for his careful and constructive reviews, which helped us improve the manuscript significantly.
We are grateful to Prof. Takaaki Noguchi for insights and fruitful discussions.
We were kindly provided with meteorite samples of Cherokee Springs, Hamlet, Harleton, Ehole, Alta'ameem, Appley Bridge, Athens, Chicora, Cynthiana, Hedjaz, Nulles, Dhajala, and Burnwell from Smithsonian National Museum of Natural History in Washington, Paragould, Ochansk, Olivenza, Chateau Renard, Soko-Banja, and Olmedilla de Alarcon from Field Museum of Natural History in Chicago, Mezo-Madaras from Naturhistorisches Museum in Vienna, and Monroe from Arizona State 
University.
Taxonomic type results presented in this work were determined, in whole or in part, using a Bus-DeMeo Taxonomy Classification Web tool by Stephen M. Slivan, developed at MIT with the support of National Science Foundation Grant 0506716 and NASA Grant NAG5-12355.
This study has utilized the JPL Small-Body Database Browser, operated at JPL, Pasadena, USA.
T.H. was supported by NASA Discovery Data Analysis Program in conducting space weathering simulation experiments in 2006-2009.
This study was supported by JSPS KAKENHI (grant nos. JP15K05277, JP17K05636, JP18K03723, and JP19H00719), by the NRF grant 2015R1D1A1A01060025 funded by the Korean government (MEST), and by the Hypervelocity Impact Facility (former facility name: the Space Plasma Laboratory), ISAS, JAXA.
\end{ack}

%

%

\end{document}